\documentclass[11pt]{article}
\pdfoutput=1

\usepackage{amsmath,amssymb}
\usepackage{epsfig}
\usepackage[numbers,compress]{natbib}
\usepackage{hyperref}
\usepackage{color}
\usepackage{slashed}
\usepackage{placeins}

\usepackage[font={small}]{caption}   

\def\SO{\textrm{SO}}

\definecolor{myred}{rgb}{0.7, 0, 0}
\definecolor{myblue}{rgb}{0, 0, 0.7}
\definecolor{mygreen}{rgb}{0.04, 0.7, 0.5}

\hypersetup{colorlinks,citecolor=myred,linkcolor=myblue,urlcolor=myblue,linktocpage=true}

 \def\be   {\begin{equation}}   \def\ee   {\end{equation}}
 \def\ba   {\begin{array}}      \def\ea   {\end{array}}
 \def\bea  {\begin{eqnarray}}   \def\eea  {\end{eqnarray}}
 \def\bean {\begin{eqnarray*}}  \def\eean {\end{eqnarray*}}
 \def\nn{\nonumber}

\newcommand{\hhref}[1]{\href{http://arxiv.org/abs/#1}{arXiv:#1}}

\numberwithin{equation}{section}

\title{
\vspace{-2cm}
\begin{flushright}
\small{CERN-PH-TH/2014-160}
\end{flushright}
\vspace{3cm}
\bf \LARGE
On the Interpretation of Top Partners Searches
\vspace{.2cm}}
\date{}
\author{
{\large Oleksii Matsedonskyi$^a$, Giuliano Panico$^{b}$, Andrea Wulzer$^{c}$}\\
[10mm]
\normalsize\itshape $^a$ Scuola Normale Superiore, Piazza dei Cavalieri 7, 56126 Pisa, Italy\\
\normalsize\itshape $^b$ CERN, Theory Division, CH-1211, Geneva 23, Switzerland\\
\normalsize\itshape $^c$ Dipartimento di Fisica e Astronomia and INFN, Sezione di Padova,\\
\normalsize\itshape Via Marzolo 8, I-35131 Padova, Italy\\
}

\begin{document}
\maketitle
\begin{abstract}
\medskip

Relatively light Top Partners are unmistakable signatures of reasonably Natural Composite Higgs models and as such they are worth searching for at the LHC. Their phenomenology is characterized by a certain amount of model-dependence, which makes the interpretation of Top Partner experimental searches not completely straightforward especially if one is willing to take also single production into account. We describe a model-independent strategy by which the interpretation is provided on the parameter space of a Simplified Model that captures the relevant features of all the explicit constructions. The Simplified Model limits are easy to interpret within explicit models, in a way that requires no recasting and no knowledge of the experimental details of the analyses.

\noindent

We illustrate the method by concrete examples, among which the searches for a charge $5/3$ Partner in same-sign dileptons and the searches for a charge $2/3$ singlet. In each case we perform a theory recasting of the available $8$~TeV Run-$1$ results and an estimate of the $13$~TeV Run-$2$ reach, also including the effect of single production for which dedicated experimental analyses are not yet available. A rough assessment of the reach of a hypothetical $100$~TeV collider is also provided.

\end{abstract}

\newpage

\tableofcontents

\newpage


\section{Introduction}

Top Partners are coloured fermions with vector-like mass associated with the Top quark. They emerge in all the New Physics scenarios where the Top--Higgs interactions, and in particular the Yukawa couplings, are generated by the mechanism of Partial Compositeness~\cite{Kaplan:1991dc}. These include the Composite Higgs (CH) scenario~\cite{gk}, as implemented in explicit five-dimensional holographic realizations~\cite{Agashe:2004rs} or in a number of four-dimensional effective parametrizations~\cite{Anastasiou:2009rv,Panico:2011pw,DeCurtis:2011yx,Matsedonskyi:2012ym}.
Top Partners are also present in other closely related scenarios such as the Little Higgs constructions, see for instance Refs.~\cite{LHTP}.

Other extensions of the SM with vector-like coloured fermions, either specifically designed to describe the CH scenario~\cite{Contino:2006nn} or not~\cite{VLQ} should be added to the list as well. However it is important to keep these models separate from the previous ones because they do not incorporate the pseudo-Nambu-Goldstone-Boson (pNGB) nature of the Higgs and describe the Top Partners by a four-dimensional renormalizable Lagrangian. Crucial features of the CH Top Partners are not captured in this approach, among which the structure of the spectrum~\cite{Panico:2011pw}, the strength of the single-production couplings~\cite{DeSimone:2012fs} and the Top Partner effects on EW Precision observables~\cite{Grojean:2013qca}. A recent attempt to address this issue is provided by the XQCUT code \cite{Barducci:2014ila}. A discussion of the latter approach in comparison with ours is postponed to the Conclusions.

Top Partners are extremely important in CH because they control the level of fine-tuning in the model~\cite{Matsedonskyi:2012ym,Panico:2012uw}: they are analog to the scalar partners of the Top in Supersymmetry. Light Top Partners, below around $2$~TeV, are unavoidably present in any ``reasonably Natural'' model which relies on less than one order of magnitude of accidental cancellation. If Natural CH is realized in Nature we should be able to discover such light Top Partners at the LHC. An exclusion would instead be an indication that the Electroweak scale is ``Unnatural'' as in the SM. In this context, alternative scenarios with non-coloured Top Partners \cite{Chacko:2005pe}, more difficult to detect, should be better investigated.

As of now, a number of Top Partner searches has been performed at the LHC using the $7$ and $8$~TeV run data~\cite{cms53,atlas051,Chatrchyan:2013uxa,ATLAS:2013ima,TheATLAScollaboration:2013oha,TheATLAScollaboration:2013sha}. More searches are expected with the $13$~TeV run, hopefully including the single-production topologies which could greatly help in extending the mass reach thanks to the large single production rates. It is time to quantify the impact of the negative $8$~TeV searches on Top Partner models and to assess the reach of the $13$~TeV ones. As described above, many models of Top Partners exist and one might be interested in performing the above study for each of them. This is not an easy task because Top Partner limits are not model-independent bounds on the mass, they depend on the strength of the coupling that controls the single production rate. They also depend on the Branching Ratios of the Top Partners in the relevant decay channels. A direct study of each given model, within which each experimental analysis should be interpreted, is too long to be performed on a case-by-case basis and must be systematized. Moreover, the comparison with the Data is difficult or impossible even within one single model if its parameter space has too many dimensions to be covered by simulations.

In order to systematize and simplify the theoretical interpretation of Top Partner searches we adopt the ``Bridge Method'', which was explicitly spelled out by one of us in Ref.~\cite{Pappadopulo:2014qza} even if it is a common implicitly adopted procedure (see e.g.~\cite{DeSimone:2012fs} in the context of Top Partners). The basic observation is that all the models describing the same kind of particles are often suited for a unified parametrization in terms of a phenomenological ``Simplified Model'', defined by a Lagrangian ${\mathcal{L}}_S$. The Lagrangian is designed to contain all and only those local interactions which emerge in the explicit models and are relevant for the experimental analyses we are interested in. The strength of the interactions and the particle masses are left as free parameters that we collectively denote as `` $\vec{c}$ '' for the present discussion. Each given explicit model, for each value of its input parameters `` $\vec{p}$ '', is reproduced by one choice $\vec{c}=\vec{c}(\vec{p})$ of the phenomenological parameters. Notice that $\vec{c}(\vec{p})$ are analytic functions which can be straightforwardly obtained by matching the explicit model Lagrangian with the Simplified one. Therefore if the experimental searches were interpreted in the Simplified Model, {\it{i.e.}} if the limits were set on the $\vec{c}$ parameters, they would be \emph{analytically} ({\it{i.e.}}, with no use of simulations and by a trivial set of numerical operations) translated in any model.

Notice that our concept of Simplified Model is rather different from the standard one of Ref.~\cite{Alves:2011wf}. In that case the Simplified Model is the description of one single signal topology while for us it is a description of all the topologies which are relevant for the particles under consideration. Furthermore, the standard prescription is to use the Simplified Model to determine the experimental signal efficiencies for the relevant topologies. Once the latter are known an automatic recasting tool can be set up for a generic model. Our procedure instead does not involve any theory recasting. The limit on the $\vec{c}$ parameters should be set directly by the experimental collaborations and the subsequent theory reinterpretation require no information on the experimental details of the analysis. The recasting which we perform in the present paper are needed only because the collaborations do not yet adopt the Simplified Model to set the limits.

\begin{figure}[t]
\begin{center}
\includegraphics[scale=0.7]{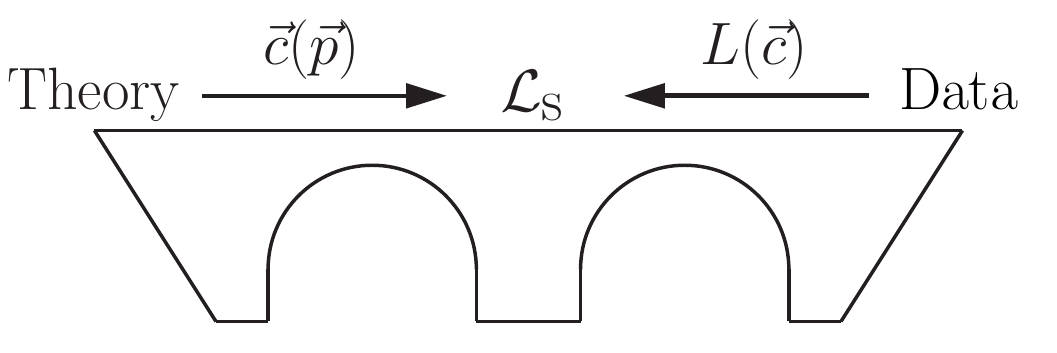}
\caption{Pictorial view of the Bridge Method.}\label{fig:uno}
\end{center}
\end{figure}

The procedure is well described by a two-span bridge depicted in Figure~\ref{fig:uno}. In the present paper we apply it to Top Partners and we focus on the ``Data'' span of the bridge. We derive the limits on the phenomenological parameters which can be inferred from the present $8$~TeV analyses and we estimate the reach of the $13$~TeV run. The ``Theory'' span will be covered in a separate publication~\cite{Matsedonskyi_inpreparation} where we discuss the impact of Top Partner searches on  concrete models. The  paper is organized as follows. In Section~\ref{sec:x53} we focus on Top Partners with electric charge $5/3$, the $X_{5/3}$, and we illustrate our method in detail in this specific example. We start by introducing the simplest possible phenomenological description of the $X_{5/3}$ and we derive the corresponding limits by reinterpreting the CMS and ATLAS searches in Refs.~\cite{cms53,atlas051} including the effect of single production.\footnote{Similar studies were performed in Refs.~\cite{DeSimone:2012fs,azatov1l}.} At a second stage we introduce a more refined treatment which takes into account that the single production vertex has not always a definite chirality and that chirality correspond to different production rates and experimental efficiencies. The general framework is introduced in Section~\ref{sec:completeframework}. It can deal with the most common Top Partners species --- namely the $T$, $B$, $X_{2/3}$, $X_{5/3}$ and $\widetilde{T}$ in the notation of Ref.~\cite{DeSimone:2012fs} --- with generic coupling chirality and also for more exotic $X_{8/3}$ \cite{Matsedonskyi:2014lla} and $Y_{-4/3}$ states. It can account for the combined effect of distinct Top Partner species contributing to the same final state and it could also be used to perform statistical combination of different channels. We apply our method to the charge $2/3$ $\widetilde{T}$ Partner and to the combined search of $B$ and $X_{5/3}$ Partners in the same-sign dilepton final state. In Section~\ref{sec:100TeV} we perform a rough assessment of the reach of a hypothetical $100\ \mathrm{TeV}$ hadronic collider.
Finally, in Section~\ref{sec:conclusions}, we present our conclusions.
After the main text, in Appendix~\ref{app:Madgraph_model} we present a \textsc{MadGraph} model designed to
simulate the Top Partners signals, while in Appendix~\ref{app:widths} we collect the analytic expressions of the
Top-Partners decay widths into SM states.


\section{The charge-5/3 partner}\label{sec:x53}

Exotic $X_{5/3}$ Partners are a generic signature of the CH scenario, where they emerge from the combined need of \mbox{SO$(4)$} custodial symmetry and of $P_{LR}$ custodial parity~\cite{Agashe:2006at}. The latter symmetries are required in order to deal with the $T$ parameter and the $Zb{\overline{b}}$ constraints respectively. Because of its origin, the $X_{5/3}$ Partner is sometimes called ``Custodian''. The $X_{5/3}$ is systematically among the lightest particles of the corresponding \mbox{SO$(4)$} multiplet. In particular it is lighter than the ordinary charge states $T$ and $B$ because, differently from the latter ones, it does not receive a positive mass shift from the mixing with the $(t_L,b_L)$ SM doublet. For this reason in many models the $X_{5/3}$ is the lightest new particle and thus the most easily accessible resonance in collider experiments. Furthermore its decay produces a rather clear signal with two energetic same-sign leptons (\emph{2ssl}). Several experimental searches of the $X_{5/3}$ have been performed by ATLAS~\cite{atlas051} and CMS~\cite{cms53} with the $7$ and $8$~TeV data. The $13$~TeV reach on this kind of particles has been also estimated~\cite{cms14}. We show below how to interpret these results in a suitable Simplified Model.

\subsection{The simplest Simplified Model}\label{sec:x53_simplest}

\begin{figure}
\centering
\includegraphics[width=.675\textwidth]{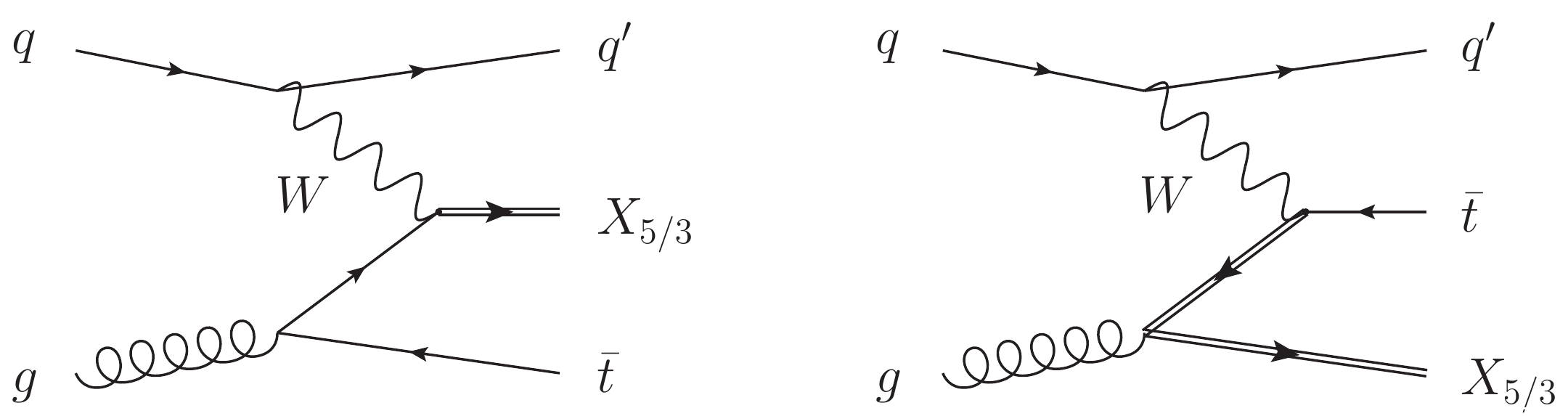}
\caption{The most relevant diagrams contributing to the $t$-associated single production of the $X_{5/3}$.}
\label{fig:spx53}
\end{figure}

Due to its peculiar properties, the $X_{5/3}$ has an extremely simple phenomenology which is captured, to a good approximation, by a simple phenomenological Lagrangian. Since it is often the lightest non-SM particle and because of its exotic charge, it typically decays to $Wt$ with unit Branching Ratio (BR). It is produced in pair by the QCD interactions or singly, through the diagrams in Fig.~\ref{fig:spx53}, by the same vertex responsible for its decay.
The simplest Simplified Lagrangian for describing the $X_{5/3}$ dynamics contains only two free parameters, the mass $M_X$ and the strength of the single-production interaction defined by
\begin{equation}\label{eq:x_53_lagr0}
{\cal L}_{5/3} =\frac{g_w}2 c_R {\overline X}_{5/3R} \slashed W  t_R + \rm{h.c.}\,,
\end{equation}
where the weak-coupling factor $g_w/2$ factor has been introduced for normalization. The only other relevant coupling is the QCD one, which however is completely fixed. We remind the reader that the $X_{5/3}$ is a color triplet like all the other Top Partners. Other interactions like the photon or the $Z$ boson couplings can be safely ignored as they give a negligible contribution to the production and are irrelevant for the decay. Basically the only non-trivial aspect of Eq.~\ref{eq:x_53_lagr0} is the choice of the chirality of the vertex. We took it Right-Handed because this is the preferred chirality in explicit models~\cite{DeSimone:2012fs} and it is not hard to understand why. One has to remember that the single-production vertex is actually the translation in Unitary Gauge of a coupling with the Goldstone boson components of the Higgs doublet and that the $X_{5/3}$ is part of one SM doublet with $7/6$ Hypercharge. Before EWSB only an \mbox{$X_{5/3}$--$H$--$t_R$} interaction is allowed, the coupling with $t_L$ is also present but it is suppressed by one insertion of the EWSB scale. It is therefore justified to ignore the Left-Handed coupling although the suppression is not so strong and, in particular corners of the parameter space, it can be overcome by numerical factors~\cite{DeSimone:2012fs}. We show in Section~\ref{sec:x_53_chiral} how to refine our treatment in order to take also this second coupling into account.

The main message of the present paper is that a Simplified Model such as the one above or its refined version described below should be employed by the experimental collaborations to interpret the $X_{5/3}$ searches. An exclusion limit or a discovery contour in the plane $(M_X,c_R)$ is immediately interpreted in any explicit Top Partner model where the two parameters can be easily computed. Here we describe a simple strategy to set limits in the $(M_X,c_R)$ plane. However possibly more complicated alternative approaches, such as for instance Matrix Element Reweighting as implemented in {\sc{MadWeight}}~\cite{Artoisenet:2010cn}, could also be considered to achieve the same goal.

We start from the basic formula for the signal yield
\begin{equation}
S = {\cal L} \sum_{n} \, \textrm{BR}_{n} \, \epsilon_{n} \, \sigma_n(M_n)\,,
\label{eq:nsignal}
\end{equation}
where $\cal L$ is the integrated luminosity and  the sum runs over the possible topologies leading to the desired final state.\footnote{A ``topology'' consists of one specific partonic production process followed by the decay in one given channel.} In the present case the sum runs over the pair and the single $X_{5/3}$ (or $\overline{X}_{5/3}$) production but in general also the production of other particles with the same signature can be taken into account as shown in Section~\ref{2P}. The $BR_n$ factors are the total Branching Ratios, accounting for the Top Partner decay ($BR(X_{5/3}\to W t)=1$ in our case) and for the subsequent decay of the heavy SM particles. When considering a \emph{2ssl} final state, $BR_{{p.p.}}\simeq 0.2$ and $BR_{{s.p.}}\simeq 0.1$ for pair and single production respectively. Finally, $\epsilon_n$ denotes the full acceptance from kinematical cuts, trigger and reconstruction efficiencies. The product $e_n=BR_n\epsilon_n$ is the total signal efficiency. We wrote Eq.~(\ref{eq:nsignal}) having in mind applications to cut-and-count experimental searches. However it is not hard to generalize it, and consequently to adapt our limit-setting strategy, to more refined shape analyses that the collaborations might decide to adopt for future searches. In this case the signal $S$ should  be promoted to the full signal shape, including normalization, and the combination $\epsilon_n\sigma_n$ should be interpreted as template shapes for the different topologies. It would be possible to parametrize the shapes semi-analytically with the same strategy discussed below for the total cross-sections $\sigma_n$.

In order to set the limits we must collect the various elements of Eq.~(\ref{eq:nsignal}), starting from the cross-sections. QCD pair production is obviously universal for all the Top Partners and independent of the single-production coupling. It only depends on the mass and it can be encapsulated in a function
\begin{equation}
\sigma_{pair}(M_X)\,,
\end{equation}
obtained by interpolating the result of Monte Carlo simulations at different mass-points for each assumed collider energy. The cross-sections are listed in Table~\ref{tab:xsecpair} for $\sqrt{s} = 8\ \mathrm{TeV}$ and \mbox{$\sqrt{s} = 13\ \mathrm{TeV}$}
center of mass energy. These results have been obtained with the HATHOR code~\cite{hathor}, which includes the QCD corrections up to NNLO, by using the MSTW2008 parton distribution functions~\cite{MSTW}. Single production is instead non-universal but it trivially scales as $c_R^2$. The reaction dominantly proceeds by the diagrams in Fig.~\ref{fig:spx53}, which consist of a forward quark splitting leading to a forward jet and to a quasi-real $W$ which scatters on the gluon producing the Top partner and the Top. Other diagrams, with the $W$ in the $s$-channel, are also included even though they give a small contribution. The cross-section can be parametrized as
\begin{equation}
\sigma_{{\textrm{sing}}}(X\overline{t})= c_R^2\, \sigma_{W^+ \overline{t}}(M_X)\,,\;\;\;\;\;\textrm{and}\;\;\;\;\;
\sigma_{{\textrm{sing}}}(\overline X t) = c_R^2\, \sigma_{W^- t}(M_X)\,,
\label{eq:x_53_prod1}
\end{equation}
for particle and anti-particle production, respectively.
At present, the coefficient functions $\sigma_{W^+ \overline{t}}(M_X)$ and $\sigma_{W^- t}(M_X)$ can be exactly computed only at LO
(for instance by using \textsc{MadGraph}~\cite{madgraph} with the dedicated model presented in Appendix~\ref{app:Madgraph_model}).
The NLO corrections, which can be significant, can only be computed with some approximated procedure.
As we will explain in Section~\ref{sec:prod_mech}, the $X_{5/3}$ single-production is closely related to the single production
of a charge $-1/3$ top partner in association with a Top quark. We can thus use the latter process, which can be implemented
in the MCFM code~\cite{mcfm}, to extract a reliable estimate of the $X_{5/3}$ single production cross-section.
The results are reported in Table~\ref{tab:xsWt} and encode the effect of the QCD interactions up to NLO,
the integration over the phase-space and the convolution with the parton distribution functions.

\begin{table}[t]
\begin{center}
\begin{tabular}{c|c|c}
\rule[-6pt]{0pt}{1.75em} & \multicolumn{2}{c}{$\sigma_{pair}$\ [fb] @ NNLO}\\
\hline
\rule[-6pt]{0pt}{1.75em}$M$\ [GeV] & $\sqrt{s} = 8$ TeV & $\sqrt{s} = 13$ TeV \\
\hline
\rule[-4pt]{0pt}{1.5em}500& 570 & $3.27 \times 10^{3}$\\
\rule[-4pt]{0pt}{1.5em}600& 169& $1.13 \times 10^{3}$ \\
\rule[-4pt]{0pt}{1.5em}700& 56.4& 442\\
\rule[-4pt]{0pt}{1.5em}800& 20.5& 190\\
\rule[-4pt]{0pt}{1.5em}900& 7.94& 87.7\\
\rule[-4pt]{0pt}{1.5em}1000& 3.21& 42.7\\
\rule[-4pt]{0pt}{1.5em}1100& 1.34& 21.7\\
\rule[-4pt]{0pt}{1.5em}1200& 0.573& 11.4\\
\end{tabular}
\hspace{1cm}
\begin{tabular}{c|c|c}
\rule[-6pt]{0pt}{1.75em} & \multicolumn{2}{c}{$\sigma_{pair}$\ [fb] @ NNLO}\\
\hline
\rule[-6pt]{0pt}{1.75em}$M$\ [GeV] & $\sqrt{s} = 8$ TeV & $\sqrt{s} = 13$ TeV \\
\hline
\rule[-4pt]{0pt}{1.5em}1300& 0.248 & 6.18\\
\rule[-4pt]{0pt}{1.5em}1400& 0.108 & 3.42\\
\rule[-4pt]{0pt}{1.5em}1500&0.047 & 1.93\\
\rule[-4pt]{0pt}{1.5em}1600&0.020 & 1.11\\
\rule[-4pt]{0pt}{1.5em}1700& --- & 0.641\\
\rule[-4pt]{0pt}{1.5em}1800& --- &0.376\\
\rule[-4pt]{0pt}{1.5em}1900& --- &0.222\\
\rule[-4pt]{0pt}{1.5em}2000& --- &0.132\\
\end{tabular}
\caption{Top partners pair production cross
section (in fb), for $\sqrt{s}=8,13$ TeV, computed at NNLO with the HATHOR code~\cite{hathor},
using the MSTW2008 parton distribution functions~\cite{MSTW}.}
\label{tab:xsecpair}
\end{center}
\end{table}

\begin{table}[t]
\centering
\begin{tabular}{ c | c | c }
\rule[-6pt]{0pt}{1.75em} & \multicolumn{2}{|c}{$\sigma_{W^+ \overline{t}}+ \sigma_{W^- t}$ $[\textrm{fb}]$ @ NLO}\\
\hline
\rule[-6pt]{0pt}{1.75em}$M$\ [GeV] & $\sqrt{s}=8$ TeV & $\sqrt{s}=13$ TeV\\
\hline
\rule[-4pt]{0pt}{1.5em}600 &  (160) 196   & (893) 1060  \\
\rule[-4pt]{0pt}{1.5em}700 &  (98.9) 124   & (613) 745  \\
\rule[-4pt]{0pt}{1.5em}800 &  (62.6) 80.3   & (431) 532  \\
\rule[-4pt]{0pt}{1.5em}900 &  (40.2) 52.8   & (308) 388  \\
\rule[-4pt]{0pt}{1.5em}1000 &  (26.2) 34.9   & (223) 285  \\
\rule[-4pt]{0pt}{1.5em}1100 &  (17.3) 23.5   & (164) 212  \\
\rule[-4pt]{0pt}{1.5em}1200 &  (11.5) 15.8 & (122) 159 \\
\rule[-4pt]{0pt}{1.5em}1300 &  (7.71) 10.8 & (90.5) 120 \\
\end{tabular}
\hspace{1cm}
\begin{tabular}{ c | c | c }
\rule[-6pt]{0pt}{1.75em} & \multicolumn{2}{|c}{$\sigma_{W^+ \overline{t}}+ \sigma_{W^- t}$ $[\textrm{fb}]$ @ NLO}\\
\hline
\rule[-6pt]{0pt}{1.75em}$M$\ [GeV] & $\sqrt{s}=8$ TeV & $\sqrt{s}=13$ TeV\\
\hline
\rule[-4pt]{0pt}{1.5em}1400 &  (5.19) 7.34 & (68.2) 91.7 \\
\rule[-4pt]{0pt}{1.5em}1500 &  (3.51) 5.04 & (51.6) 70.6 \\
\rule[-4pt]{0pt}{1.5em}1600 &  (2.39) 3.48 & (39.3) 54.1 \\
\rule[-4pt]{0pt}{1.5em}1700 & ---        & (30.2) 42.0  \\
\rule[-4pt]{0pt}{1.5em}1800 & ---        & (23.2) 32.4  \\
\rule[-4pt]{0pt}{1.5em}1900 & ---        & (17.9) 25.2  \\
\rule[-4pt]{0pt}{1.5em}2000 & ---        & (13.9) 19.8  \\
\multicolumn{3}{c}{\rule[-4pt]{0pt}{1.5em}}
\end{tabular}
\caption{NLO single production cross sections for the $Wt$ fusion for a unit coupling,
at $\sqrt{s}=8,13$ TeV (the LO values are in brackets) computed with
MCFM \cite{mcfm} by considering the closely related process of single production of
a charge $-1/3$ Top Partner $pp \rightarrow Bt$ (see main text for more details).
The results were obtained by using the MSTW2008 parton distribution functions.}
\label{tab:xsWt}
\end{table}

Now that the cross-sections are known, all what is left to compute are the acceptance factors $\epsilon_{p.p}$ and $\epsilon_{s.p.}$. The important point is that the latter factors only depend on the kinematical distributions of the pair and single production topologies and not on their normalization. As such they do depend on the resonance mass but not on the coupling which merely rescales the total rate. The efficiencies at each mass point can thus be obtained by two template Monte Carlo simulations, one for the pair and the other for the single production topologies.\footnote{The single production of the $X_{5/3}$ and of its anti-particle can be treated as a single topology because the efficiencies are charge-symmetric.} Ideally, the coupling could affect the kinematical distributions and consequently the efficiencies through the finite resonance decay width. However the effect is negligible for narrow enough Partners. Below we estimate the efficiencies and we draw exclusion limits based on the Run-1 LHC analyses at $8$~TeV and on projections for Run-2.

\subsection{Efficiencies and bounds}\label{sec:efficiency}

After defining our simplified set-up, we now show how it can be used to interpret the LHC results.
As a first step we take into account the $8$~TeV LHC run to derive some bounds on the mass of the
exotic $X_{5/3}$ resonance. Afterwards we perform an exploratory analysis of the Run-2 LHC reach.
We postpone to Section~\ref{sec:100TeV} an analysis of the reach of a hypothetical $100\ \mathrm{TeV}$
hadron collider.

Our starting point are the recent experimental analyses performed by ATLAS~\cite{atlas051} and CMS~\cite{cms53} searching for \emph{2ssl} final states, the cleanest signal of a charge-$5/3$ Top Partners.
CMS provides an interpretation of the limits for an $X_{5/3}$ signal. On the other hand, ATLAS assumes a charge $-1/3$ $B$ partner, which,
as we will explain in the following sections, has a phenomenology very similar to the $X_{5/3}$.
Both searches consider Top Partner QCD pair production only but, in fact, the analyses are potentially also sensitive to $X_{5/3}$ single production. The simplicity of these analyses, which are bases on a cut-and-count strategy, allows us to perform a straightforward recast of the results, as described below.

The CMS search~\cite{cms53} is based on $19.6\,\mathrm{fb}^{-1}$ of collected data, it looks for an excess of events containing \emph{2ssl} ($e$ or $\mu$, including those from $\tau$ decays) and at least $N_{con}=5$ additional constituents, \emph{i.e.} other leptons or parton-level jets. A dedicated technique is used to reconstruct top quarks and $W$-bosons from their decay products if the latter are highly boosted. The candidate leptons and jets are required to satisfy isolation criteria, minimum $p_\bot$ and $\eta$ cuts and the invariant mass of the leptons pairs must be away from  the $Z$ peak to further suppress the $WZ$ and $ZZ$ background. On top of this, the sum of the transverse momenta of the particles in the event must be larger than $900$~GeV. The search did not find any significant excess and put a lower limit of $770$~GeV
on the mass of charge $5/3$ states at the $95\%$ confidence level.\footnote{Notice that a more recent version of the CMS analysis~\cite{cms53_new} quotes a slightly higher bound ($M_{X} \geq 800\ \mathrm{GeV}$). For our recast, however, we will stick to the earlier version because the latter, unlike the former one, reports separately the cut efficiencies which can be used to check the reliability of our recast. } This bound corresponds to an upper limit $S_{exc}^{\text{CMS}}\simeq 12$ on the signal events passing the selection criteria. Notice that the analysis assumes that the $W$-mediated interactions of the $X_{5/3}$ with the top quark are vector-like, \emph{i.e.} that the resonance couples with equal strength to the Left- and Right-handed top components. As we saw before, this does not coincide with the expected coupling pattern with purely chiral interactions and leads to a mild shift in the efficiencies and thus in the resulting mass limit.

Though the bound on the pair production signal cross section obtained by the CMS analysis is stronger than the one of ATLAS~\cite{atlas051} the latter one turns out to be more sensitive to the single production topology due to the different selection cuts. In particular the ATLAS analysis applies a much milder cut on the total number of constituents (only two jets are required rather than $5$ constituents) and this makes the cut acceptance higher than for the CMS one. Indeed singly produced resonances lead to at most $5$ parton-level jets, one of which is very forward and has a low $p_\bot$. Loosing one of those, especially the forward one, is extremely likely.  Apart from exactly two same sign leptons and two additional jets, the ATLAS search requires at least one $b$-tagged jet. Like in the CMS one, the jets and leptons candidates must satisfy isolation criteria, minimum $p_\bot$ and $\eta$ cuts and the invariant mass of the lepton pair must be away from the $Z$ mass. In addition to this, there should be a missing transverse energy $E_{\text T}^{\text{miss}} > 40$~GeV and the scalar sum of the $p_\bot$'s of all the jets and leptons must be greater than $650$~GeV. The search is based on $14.3\ \mathrm{fb}^{-1}$ of integrated luminosity and provides, given the observed cross-section limit, an upper bound $S_{exc}^{\text{ATLAS}}\simeq 13$ on the number of signal events. The interpretation is provided for a $B$ bottom-like excited state, which is assumed to couple only to the Left-handed Top component.

\subsubsection{Event selection efficiency}

The production cross sections of the $X_{5/3}$ resonance have been already discussed. The only missing ingredients for our analysis are thus the cut acceptances. To compute them we used our {\sc{MadGraph}}~\cite{madgraph} model \cite{HEPMDB}, described below in Section~\ref{sec:completeframework}, which contains the $X_{5/3}$ resonance and its coupling to the top quark in Eq.~(\ref{eq:x_53_lagr0}). The latter coupling is responsible for both single production and for the decay. We generated the events by using {\sc{MadGraph}} and we used PYTHIA~\cite{Sjostrand:2006za} to include parton showering effects. Jet clustering and lepton isolation criteria were performed on the showered events and the kinematical cuts were applied on the resulting reconstructed objects. The $b$-tagging (needed for the recast of the ATLAS search), lepton reconstruction and trigger efficiencies were assumed to be independent of the kinematics and were taken into account through universal reweighting factors reported in the experimental papers. The efficiency for leptonically-decaying Tau's was tuned in order to maximize the agreement with the ATLAS and CMS efficiencies over the whole Top Partner mass range. The boosted $W$ and top reconstruction algorithm (needed for the CMS search) was also applied on the showered events. We estimated the reliability of our recast by reproducing the efficiencies reported in the ATLAS and CMS analyses within their signal hypothesis, namely a $B$ coupled to the Left-handed Top in the case of ATLAS and an $X_{5/3}$ with vector-like coupling for CMS. We also reproduced the single-production efficiency for the ATLAS search derived in Ref.~\cite{Ortiz:2014iza}.

The signal efficiencies obtained by our recast are reported in the Tables~\ref{tab:2sslcms8} and~\ref{tab:2sslatlas8} for the single and pair production topologies. The ones relevant for the present discussion, derived assuming purely Right-Handed couplings, are reported in the first column of the tables. The second one is described and employed in  Section~\ref{sec:x_53_chiral}.
\begin{table}[t]
\begin{center}
\begin{tabular}{ c | c | c | c  }
\multicolumn{4}{c}{\rule[-5pt]{0pt}{1.em}CMS,\ \  single prod. eff. [$\%$]} \\
\hline
\!$M\ \textrm{[GeV]}$\!
& $\begin{array}{c} \rule[-4pt]{0pt}{1.75em}Q=\tfrac{5}{3} \\ \rule[-6pt]{0pt}{1.65em}\textrm{right} \end{array}$
& $\begin{array}{c} \rule[-4pt]{0pt}{1.75em}Q=\tfrac{5}{3} \\ \rule[-6pt]{0pt}{1.65em}\textrm{left} \end{array}$
&  $\begin{array}{c} \rule[-4pt]{0pt}{1.75em}Q=\tfrac{5}{3} \\   \rule[-6pt]{0pt}{1.65em}\textrm{left=right~\cite{azatov1l}}\!\!\end{array}$  \\
\hline
\rule{0pt}{1.2em}700 & 0.185 & 0.165  & 0.201 \\
\rule{0pt}{1.em}800 & 0.269  & 0.210 & 0.266 \\
\rule{0pt}{1.em}900 & 0.308  & 0.237 & 0.312 \\
\end{tabular}
\hfill
\begin{tabular}{  c | c | c | c  }
\multicolumn{4}{c}{\rule[-5pt]{0pt}{1.em}CMS,\ \ pair prod. eff. [$\%$]} \\
\hline
\!$M\ \textrm{[GeV]}$\!
& $\begin{array}{c} \rule[-4pt]{0pt}{1.75em}Q=\tfrac{5}{3} \\ \rule[-6pt]{0pt}{1.65em}\textrm{right} \end{array}$
& $\begin{array}{c} \rule[-4pt]{0pt}{1.75em}Q=\tfrac{5}{3} \\ \rule[-6pt]{0pt}{1.65em}\textrm{left} \end{array}$
&  $\begin{array}{c} \rule[-4pt]{0pt}{1.75em}Q=\tfrac{5}{3} \\   \rule[-6pt]{0pt}{1.65em}\textrm{left=right~\cite{cms53}}\!\!\end{array}$  \\
\hline
\rule{0pt}{1.2em}700 & 2.27  & 1.66 & 1.85 \\
\rule{0pt}{1.em}800 & 2.64  & 1.95 & 2.33 \\
\rule{0pt}{1.em}900 & 2.85  & 2.19 & 2.57 \\
\end{tabular}\end{center}
\caption{Total signal efficiency $e$ for the CMS analysis~\cite{cms53} for a single- (left table)
and pair-produced (right table) charge-$5/3$ top partner.
The results are given for purely right- and purely left-handed couplings.
The last columns show the values of the efficiencies extracted from
the Refs.~\cite{cms53,azatov1l}.}
\label{tab:2sslcms8}
\end{table}
\begin{table}[t]
\begin{center}
\begin{tabular}{ c | c | c}
\multicolumn{3}{c}{\rule[-5pt]{0pt}{1.em}ATLAS,\ \ single prod. eff. [$\%$]} \\
\hline
$M\ \textrm{[GeV]}$
& $\begin{array}{c} \rule[-4pt]{0pt}{1.75em}Q=\tfrac{5}{3} \\ \rule[-6pt]{0pt}{1.65em}\textrm{right} \end{array}$
& $\begin{array}{c} \rule[-4pt]{0pt}{1.75em}Q=\tfrac{5}{3} \\ \rule[-6pt]{0pt}{1.65em}\textrm{left} \end{array}$   \\
\hline
 \rule{0pt}{1.2em}700  & 1.14  & 0.952  \\
 \rule{0pt}{1.em}800  & 1.26  & 1.01  \\
 \rule{0pt}{1.em}900  &  1.31  & 1.10  \\
 \rule{0pt}{1.em}1000 &  1.23 & 1.09  \\
 \rule{0pt}{1.em}1100 &  1.26 & 1.13  \\
 \rule{0pt}{1.em}1200 &  1.25 & 1.19  \\
\end{tabular}
\hspace{1.2cm}
\begin{tabular}{ c | c | c | c }
\multicolumn{4}{c}{\rule[-5pt]{0pt}{1.em}ATLAS,\ \  pair prod. eff. [$\%$]} \\
\hline
$M\ \textrm{[GeV]}$
& $\begin{array}{c} \rule[-4pt]{0pt}{1.75em}Q=\tfrac{5}{3} \\ \rule[-6pt]{0pt}{1.65em}\textrm{right} \end{array}$
& $\begin{array}{c} \rule[-4pt]{0pt}{1.75em}Q=\tfrac{5}{3} \\ \rule[-6pt]{0pt}{1.65em}\textrm{left} \end{array}$
&  $\begin{array}{c} \rule[-4pt]{0pt}{1.75em}Q=-\tfrac{1}{3}\ \ (b^{\prime}) \\   \rule[-6pt]{0pt}{1.65em}\textrm{left~\cite{atlas051} } \end{array}$  \\
\hline
\rule{0pt}{1.2em}700  & 2.17   & 1.87 & 1.84 \\
\rule{0pt}{1.em}800  & 2.23   & 1.95 & 2.03 \\
\rule{0pt}{1.em}900  & 2.22   & 2.00 & 2.06 \\
\rule{0pt}{1.em}1000 & 2.23  & 2.03 & -- \\
\rule{0pt}{1.em}1100 & 2.24  & 2.07 & -- \\
\rule{0pt}{1.em}1200 & 2.23  & 2.06 & -- \\
\end{tabular}\end{center}
\caption{Total signal efficiency $e$ for the ATLAS analysis~\cite{atlas051} for a single- (left table)
and pair-produced (right table) charge-$5/3$ top partner.
The results are given for purely right- and purely left-handed couplings.
The last column of the right table shows the efficiencies extracted from
the Ref.~\cite{atlas051} for the case of a fourth generation $b^{\prime}$ quark.}
\label{tab:2sslatlas8}
\end{table}
As expected, the CMS analysis has a very strong preference for events coming from pair produced resonances. Indeed the signal efficiency for single production is extremely low, an order of magnitude smaller than the pair-production one. The situation is different for the ATLAS analysis. In this case the signal efficiency for a singly-produced resonance is only a factor $2$ smaller than the one for pair production and including both production modes in the analysis can lead to a sizeable enhancement of the bounds.

\subsubsection{Exclusions}

We now present the result of our recast in terms of the relevant parameters of the simplified model, namely the resonance mass $M_{X}$ and the single-production coupling $c_R$. As a function of these parameters we can compute the number of signal events expected for the CMS and ATLAS analyses and compare them with the experimental bounds $S_{exc}^{\text{CMS}}$ and $S_{exc}^{\text{ATLAS}}$. The exclusion bounds on $M_{X}$ are shown in Fig.~\ref{fig:2sslexclgen} as a function of the coupling $c_R$.
\begin{figure}[t]
\centering
\includegraphics[width=0.475\textwidth]{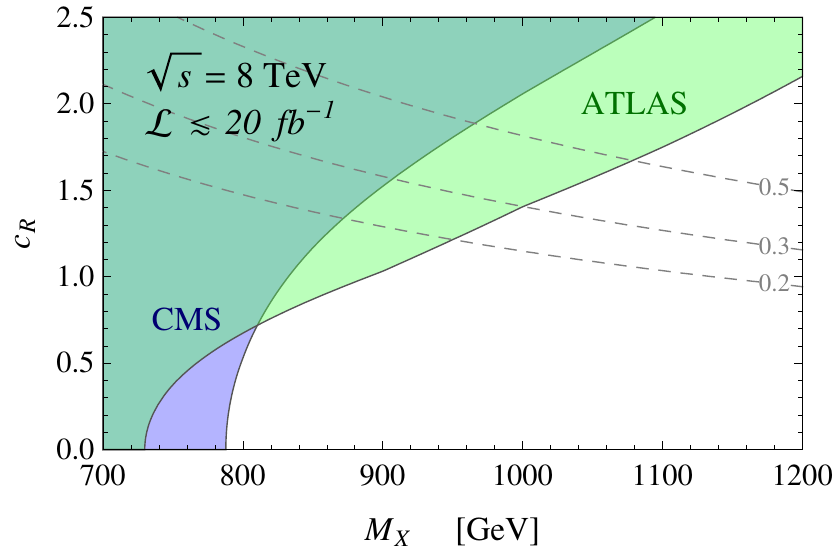}
\caption{Bounds on the mass of a charge-$5/3$ state, decaying exclusively
to $W t$ as a function of the single-production coupling $c_R$.
The $c_R$ coupling is assumed to be the only relevant coupling of the resonance
with the SM quarks. The green and blue shaded regions correspond to the ATLAS and CMS bounds
respectively. The dashed gray lines show the contours with $\Gamma_X/M_X = 0.2, 0.3, 0.5$.}
\label{fig:2sslexcl_x53_R}
\end{figure}

As expected, for low values of the coupling $c_R \lesssim 0.7$, when pair production dominates, the CMS analysis has a better sensitivity than the ATLAS one. Notice that the bound at very small values of the coupling, $M_{X} > 790\ \mathrm{GeV}$, does not coincide with the limit quoted by CMS ($M_{X} > 770\ \mathrm{GeV}$) because the latter assumes a vector-like coupling rather than a Right-Handed one.  As the $c_R$ coupling gets larger the CMS bound  only mildly increases due to the small single production acceptance. For higher values of the coupling $c_R \gtrsim 0.7$, thanks to the sizable contribution coming from single production, the ATLAS analysis becomes more sensitive than the CMS one and leads to a bound that steeply increases with the size of the coupling. Contours of fixed $X_{5/3}$ width over mass ratio are also shown in the plot. We notice that the resonance is typically narrow for $c_R \lesssim 1$ while for larger couplings, especially for $M_X \gtrsim 1\ \mathrm{TeV}$, the width becomes significant and it could start affecting the bounds. For simplicity
we did not include those effects in our analysis.

After the recast of the current experimental searches, we want to use our simplified approach to estimate the future reach of the $13\ \mathrm{TeV}$ LHC Run-$2$. As far as pair production is concerned, a robust starting point is provided by Ref.~\cite{cms14}, where the \emph{2ssl} channel is analysed in some detail. The pair production efficiency in the relevant mass region ($1.2\ \mathrm{TeV} \lesssim M_X \lesssim 2\ \mathrm{TeV}$) is found to depend only mildly on $M_X$ and it varies in the range $1.3 \% - 1.7 \%$, we thus assume a uniform efficiency of $1.5 \%$ in our analysis. We also ignore the fact that a non-chiral coupling was employed in Ref.~\cite{cms14}. No study is available for single production, and furthermore we have seen that the single production efficiency strongly depends on the selection strategy. Not having any hint on how the single production search will be performed at Run-$2$ we consider $3$ possible scenarios. In the first one we assume that the single production efficiency will be much lower than the pair-production one, namely $e_{s.p.} = 0.1\, e_{p.p.}$, which is what happens for the $8$~TeV CMS search. This pessimistic scenario is unrealistic, but it clearly shows the need of a dedicated analysis for single production. The second scenario assumes $e_{s.p.} = 0.5\, e_{p.p.}$ in analogy with the $8$~TeV ATLAS search. As a third possibility we consider the case $e_{s.p.} = e_{p.p.}$ which believe to be realistically achievable by a dedicated search. The number of expected background event, with the cuts of Ref.~\cite{cms14}, is $B\simeq10$ for \mbox{$300$fb$^{-1}$} of integrated luminosity. By rescaling we easily obtain the background for different luminosities and thus we estimate the minimal number of signal events needed for exclusion. We take $S_{exc.}=3\sqrt{B}$ for $B>1$ and $S_{exc.}=3$ if $B<1$. This of course relies on the assumption that the background cross-section will be approximately the same also for the single production dedicated analyses.

\begin{figure}[t]
\centering
\includegraphics[width=0.475\textwidth]{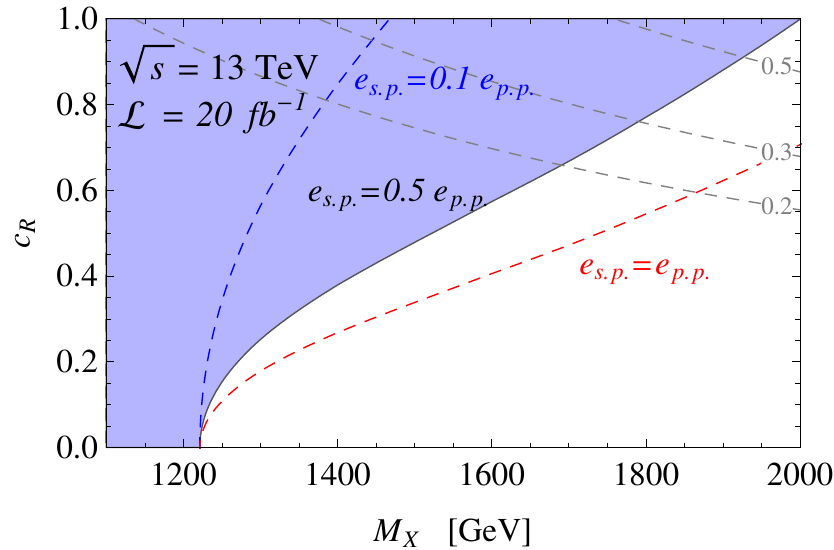}
\hfill
\includegraphics[width=0.475\textwidth]{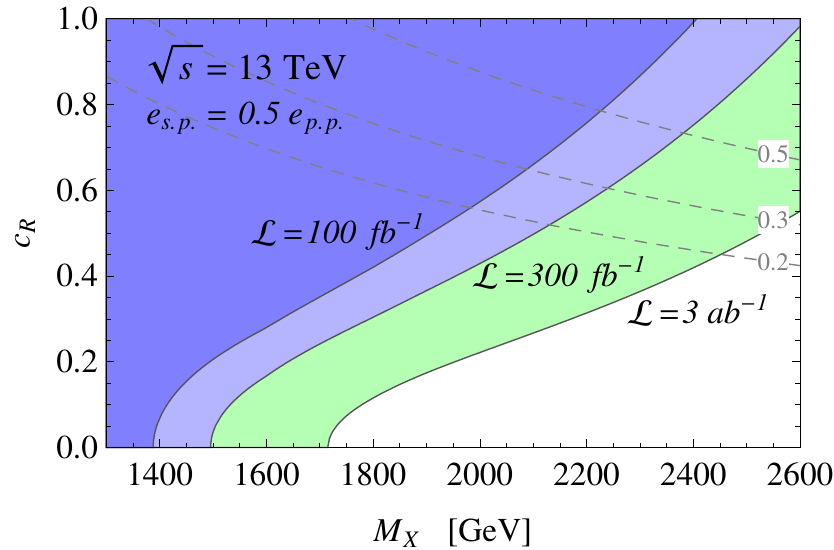}
\caption{Estimated exclusion reach for the mass of a charge-$5/3$ state decaying exclusively
to $W t$ as a function of the $c_R$ coupling. To obtain the excluded regions we assumed
$\sqrt{s} = 13\ \mathrm{TeV}$ collider energy and
$L = 20\ \mathrm{fb}^{-1}$ integrated
luminosity (left panel) and $L = 100, 300, 3000\ \mathrm{fb}^{-1}$
integrated luminosity (right panel). The dashed gray lines show the contours with $\Gamma_X/M_X = 0.2, 0.3, 0.5$.}
\label{fig:2sslexclgen_13}
\end{figure}

The results are reported in Fig.~\ref{fig:2sslexclgen_13}. We see that \mbox{$20$ fb$^{-1}$} of integrated luminosity could put, in the absence of a signal, a coupling-independent limit $M_X>1.2$~TeV from QCD pair production. The limit can reach $2$~TeV for sizeable single production coupling strength. The figure also shows, on the right panel, the projections for \mbox{$100$ fb$^{-1}$} ({\it{i.e.}} the final luminosity goal of Run-$2$), for \mbox{$300$ fb$^{-1}$} and \mbox{$3000$ fb$^{-1}$}.

\subsection{A slight refinement}\label{sec:x_53_chiral}

In most cases the Simplest Simplified Model provides an accurate description of the $X_{5/3}$ phenomenology, however there are corners of the parameter space of explicit models where other effects should be taken into account. The most relevant one is the presence of a Left-Handed single production coupling, which leads us to turn Eq.~(\ref{eq:x_53_lagr0}) into
\begin{equation}
{\cal L}_{5/3} = \frac{g_w}2  c_R \, \overline X_{5/3 R} \slashed W  t_R
+ \frac{g_w}2  c_L \, \overline X_{5/3 L} \slashed W  t_L  + \text{h.c.}\,.
\label{eq:x_53_lagr1}
\end{equation}
As explained above, $c_L$ is structurally suppressed with respect to $c_R$, however it can become comparable or even larger than the latter in some cases. Below we show how this new parameter can be taken into account by setting limits in the $3$-dimensional parameter space $(m_X,c_R,c_L)$ of this more refined Simplified Model. This also allows us to assess the accuracy of the Simplest Simplified Model and the robustness of the limits derived in the previous Section.

\begin{figure}
\centering
  \includegraphics[width=0.48\textwidth]{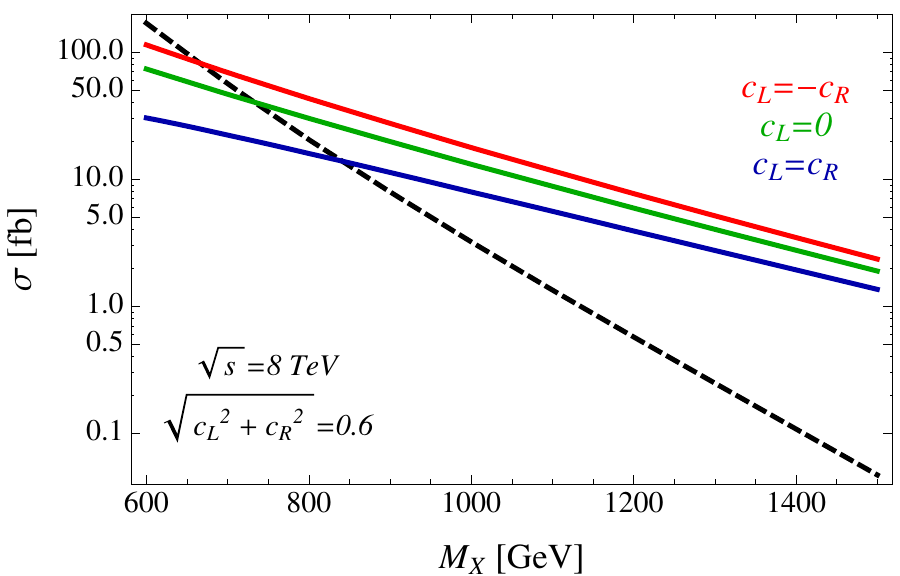}   \hfill
  \includegraphics[width=0.48\textwidth]{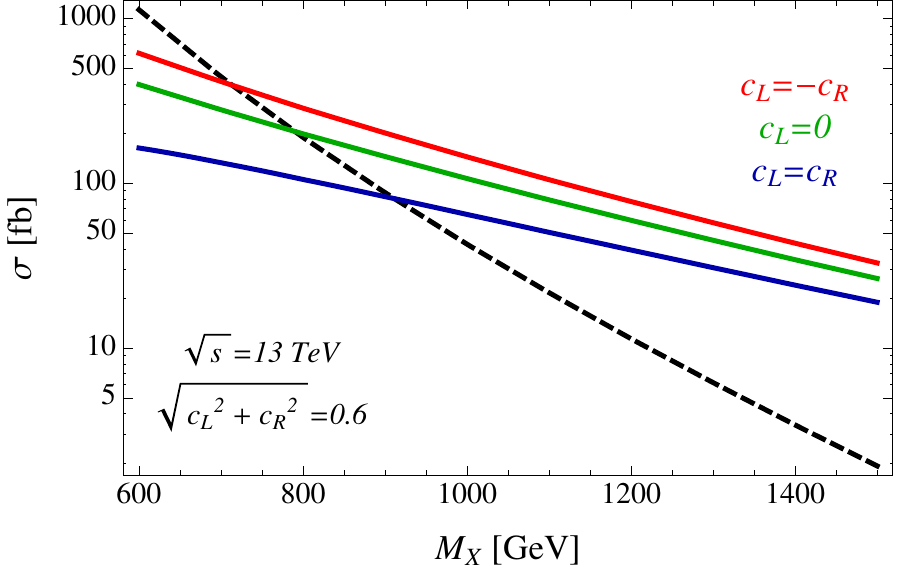} \\
  \caption{Cross sections of $X_{5/3}$ pair (black dashed) and single production for $\sqrt{c_L^2+c_R^2}=0.6$ and $c_L=c_R$ (blue), $c_L=0$ or equivalently $c_R=0$ (green) and $c_L=-c_R$ (red),  for $\sqrt s = 8$~TeV (left panel) and $\sqrt s = 13$~TeV (right panel).}
  \label{fig:xs_53}
\end{figure}

The first effect of the new coupling is to modify the theoretical prediction of the single-production cross-section. The Feynman amplitude of the process, in Figure~\ref{fig:spx53}, is now the sum of two terms, proportional to $c_R$ and $c_L$, respectively. The cross-section is thus the sum of three terms scaling as $c_R^2$, $c_L^2$ and $c_Lc_R$ from the interference. Given that the QCD interactions are Left--Right symmetric, the $c_R^2$ and $c_L^2$ coefficients are identical and can be parametrized by the same coefficient functions $\sigma_{W^+\overline{t}}(M_X)$ and ${ \sigma}_{W^-t}(M_X)$ introduced in Eq.~(\ref{eq:x_53_prod1}) for $X_{5/3}$ and ${\overline{X}}_{5/3}$, respectively. The interference term is suppressed by the fact that it must vanish in the limit of zero Top mass because in that limit the chirality of the Top quark or anti-quark produced in association with the resonance becomes a physical observable and the two couplings can not interfere. Since the center-of-mass energy of the $W^*$--gluon collision that produces the resonance is approximately set by the production threshold $m_t+M_X$ a suppression of order $m_t/(m_t+M_X)$ of the interference is expected. We thus find convenient to parametrize
\begin{eqnarray}
\sigma_{{\textrm{sing}}}(X\overline{t}) = \left(c_R^2+c_L^2\right)\sigma_{W^+ \overline{t}}(M_X) +  c_R\, c_L \left( {m_t \over M_X+m_t} \right) \sigma^{\prime}_{W^+ \overline{t}}(m_X) \,,\nonumber \\
\sigma_{{\textrm{sing}}}(\overline X t)=\left(c_R^2+c_L^2\right) \sigma_{W^- t}(M_X) +  c_R\, c_L \left( {m_t \over M_X+m_t} \right) \sigma^{\prime}_{W^- t}(M_X) \,.
\label{eq:prod1}
\end{eqnarray}
The interference coefficient functions $\sigma^{\prime}_{W^+ \overline{t}}(M_X)$ and $\sigma^{\prime}_{W^- t}(M_X)$ can be extracted at each mass-point by a pair of Monte Carlo simulations at $\{c_R=c,\,c_L=0\}$ and $c_R=c_L=c/\sqrt{2}$. However the MCFM code does not allow to change the coupling chirality and we must content ourselves with a LO estimate done with {\sc{MadGraph}}~\cite{madgraph}. It turns out that $\sigma^{\prime}_{Vt}(M_X)$ is very well approximated, both at $8$ and $13$~TeV collider energy, by
\be\label{eq:intefr}
\sigma^{\prime}_{W^+ \overline{t}}(M_X)  \simeq - 5.2 \, \sigma_{W^+ \overline{t}}(M_X)\,.
\ee
The same holds for the charge conjugated process. We checked that Eq.~(\ref{eq:intefr}) holds up to few percent corrections in the mass range $600\ \mathrm{GeV} \leq M_X \leq 2000\ \mathrm{GeV}$. Because of this numerical enhancement the contribution of the interference to the total rate can be considerable. As shown in Fig.~\ref{fig:xs_53} it is of order unity in the relevant mass range.

\begin{figure}
\centering
\includegraphics[width=0.49\textwidth]{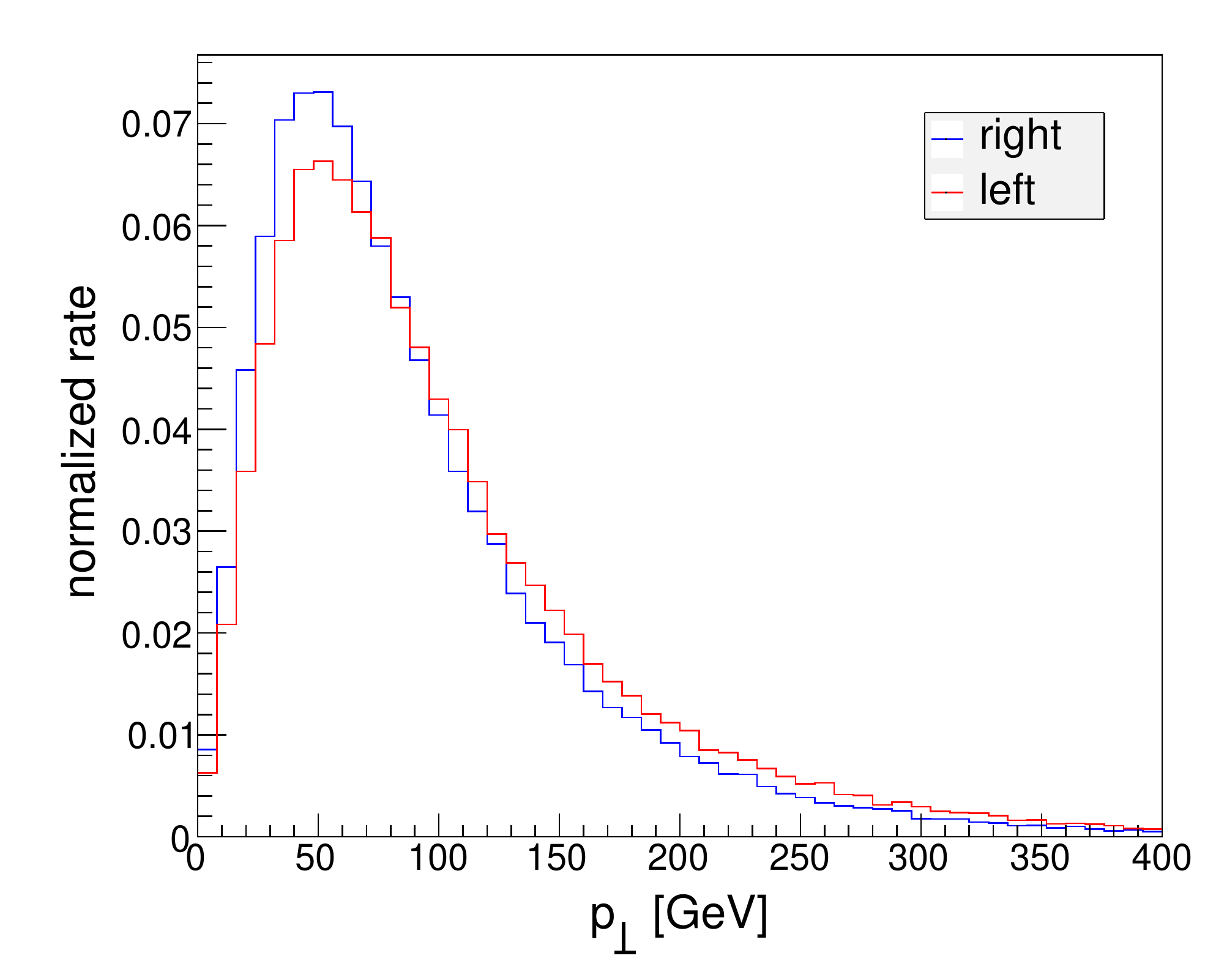}   \hfill
\includegraphics[width=0.49\textwidth]{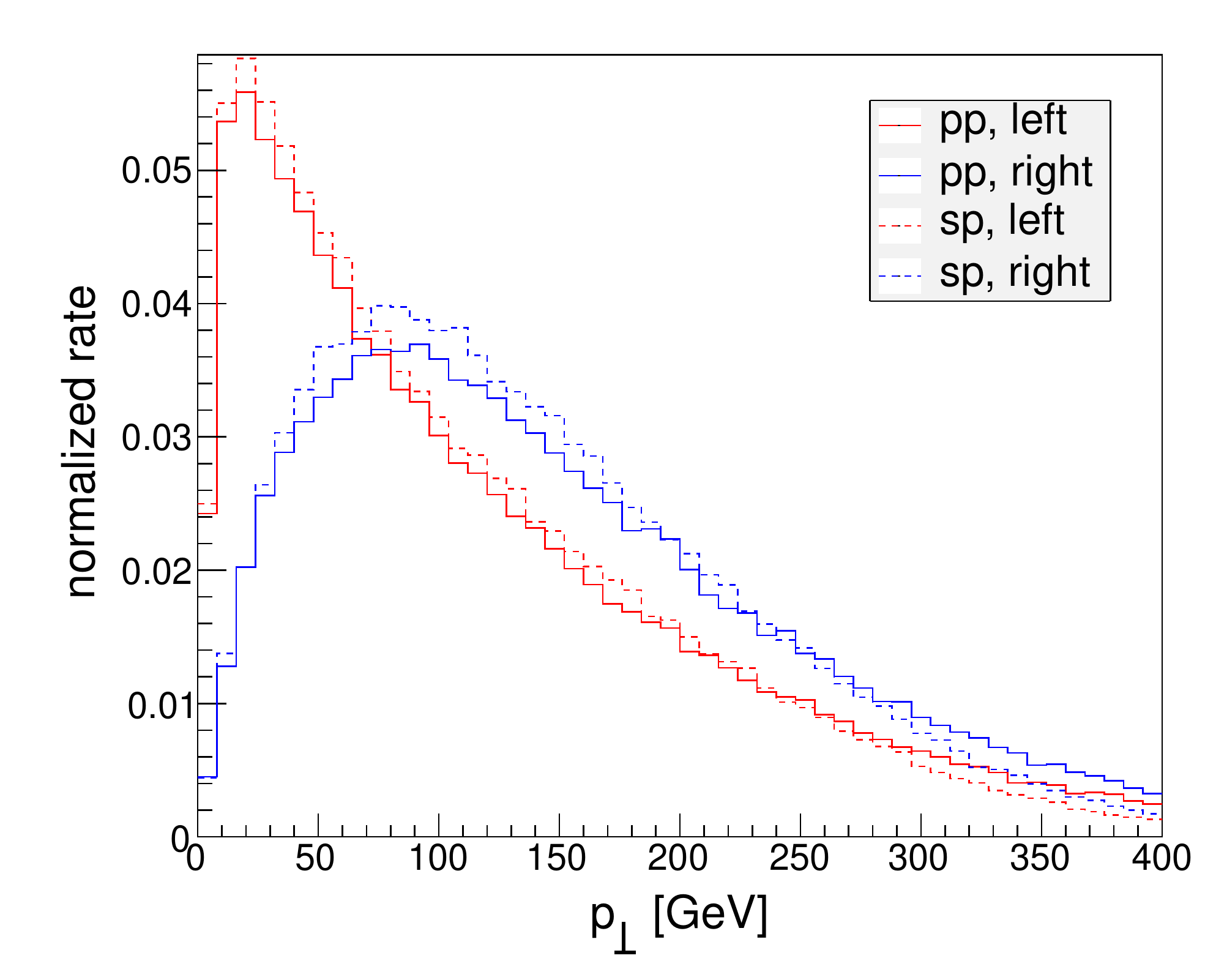} \\
\caption{On the left panel: $p_\bot$ distribution of the Bottom quark from the associated Top quark in single production.
On the right panel: $p_\bot$ distributions of the same-sign leptons in the cases of $X_{5/3}$ pair (solid lines) and single (dashed lines) production. In all the plots the red lines correspond to the scenario
with purely Left-Handed coupling to the top quark, while the blue lines correspond to
purely Right-Handed coupling. The mass of the $X_{5/3}$ has been fixed to $800$~GeV
and the collider energy to $\sqrt s = 8$~TeV.}
\label{fig:pt_e}
\end{figure}

The coupling chirality also affects the kinematical distributions of the final state objects --- namely leptons, $E_T^{\textrm{miss}}$, jets and $b$-jets --- employed for event selection and thus it modifies the signal efficiencies. This second effect turns out to be much less relevant than the modification of the cross-section and it could be safely neglected. However it is interesting to see how it can be taken into account with our method. The kinematical distributions are distorted by two distinct effects. First, by the chirality of the Top quark or anti-quark produced in association with the resonance in the single production mode. The chirality affects the helicity of the associated Top, which in turn determines the decay products distributions because of Spin Correlations. However the effect is marginal because the associated Top is mainly produced at low velocity and thus its helicity has a small impact on the final states. This is confirmed by the left panel of Figure~\ref{fig:pt_e} where we show the $p_\bot$ distribution of the bottom from the associated Top decay. Those of the additional decay products, namely the two light jets, have identical shapes for the two coupling chiralities and thus they are not shown in the plot. The second effect has a similar physical origin, but it is quantitatively more relevant. It has to do with the chirality of the Top from the resonance decay. When the latter is heavy the Top is considerably boosted and Spin Correlations affect the distributions of its products in a significant way, as shown in the right panel of Figure~\ref{fig:pt_e}. The Right-Handed coupling tends to produce more energetic leptons, making easier for this configuration to pass the acceptance cuts on the two same-sign leptons $p_\bot$. We can take this effect into account by introducing a mild dependence of the efficiencies on the couplings, namely
\begin{equation}
e_n=\frac{c_L^2}{c_L^2+c_R^2} e_n^L+\frac{c_R^2}{c_L^2+c_R^2} e_n^R\,,
\label{reff}
\end{equation}
where $e_n^{L,R}$ are the efficiencies for purely Left- and purely Right-Handed couplings. The parametrization above, whose accuracy has been checked both for the single and for the pair production mode, follows from the fact that the fraction of Left- and Right-Handed Top quarks from the $X_{5/3}\rightarrow Wt$ decay is controlled by the factors $c_L^2/(c_L^2+c_R^2)$ and $c_R^2/(c_L^2+c_R^2)$, respectively. The Left- and Right-Handed efficiencies are reported in Tables~\ref{tab:2sslcms8} and~\ref{tab:2sslatlas8} for the ATLAS and CMS \emph{2ssl} $8$~TeV searches. We derived them by simulations as described in Sect.~\ref{sec:efficiency}.

\begin{figure}[t]
\centering
\includegraphics[width=0.475\textwidth]{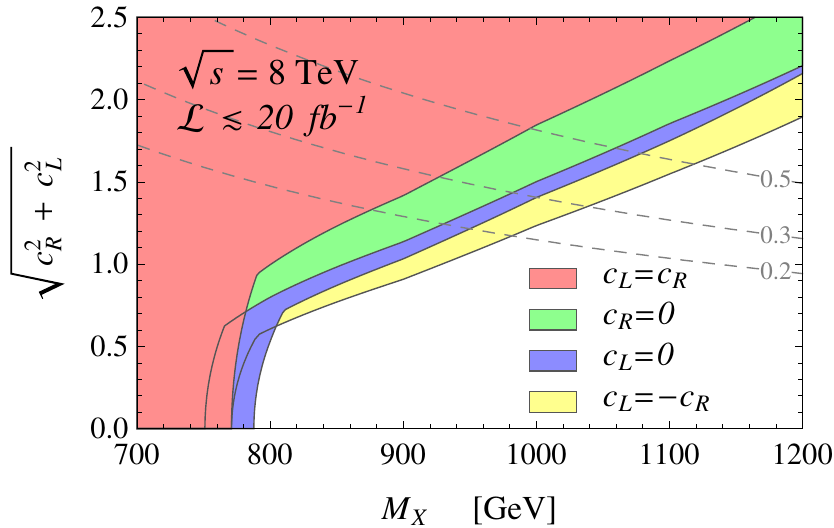}
\hfill
\includegraphics[width=0.475\textwidth]{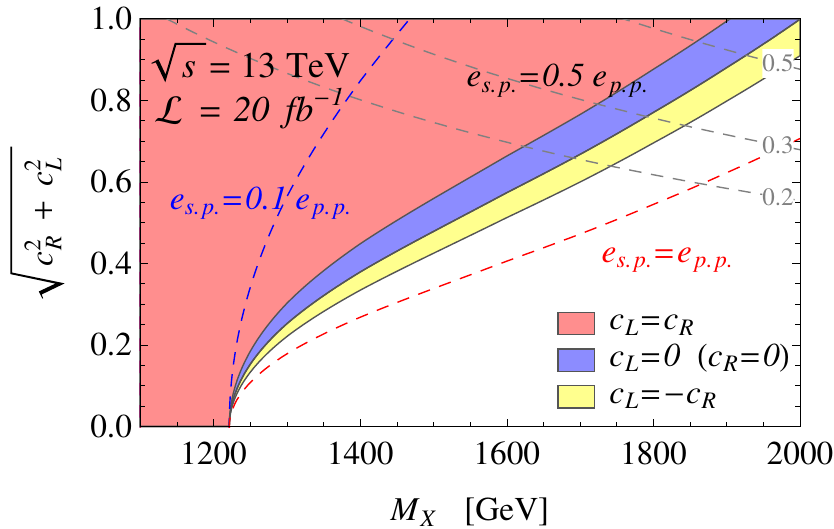}
\caption{Bounds on the mass of charge-$5/3$ resonance, decaying exclusively
to $W t$, for different combinations of the left ($c_L$) and right ($c_R$) couplings to the
top quark.
The left panel shows the bound for the $8\ \mathrm{TeV}$ LHC, while the right panel shows the
expected bounds for $13\ \mathrm{TeV}$ collider energy with $L = 20\ \mathrm{fb}^{-1}$ integrated luminosity.
The dashed gray lines show the contours with $\Gamma_X/M_X = 0.2, 0.3, 0.5$.}
\label{fig:2sslexclgen}
\end{figure}

As anticipated, the difference between the Left- and Right-Handed efficiencies is rather mild. The corrections introduced by Eq.~(\ref{reff}), relative to the case of flat efficiencies $e_n=e_n^L$ are below around $30\%$ for ATLAS and $20\%$ for CMS and could be safely ignored. However for completeness we take them into account in the final $8$~TeV exclusion plot reported in the left panel of Fig.~\ref{fig:2sslexclgen}, where the limit is set in the $\sqrt{c_L^2+c_R^2}$ versus mass plane. By comparing with our previous result in Fig.~\ref{fig:2sslexcl_x53_R}, which corresponds to the $c_R=0$ contour, we see that chirality effects, due to the change in the cross-section, can be rather significant. Because of the enhanced interference the mass limit can vary by around $100$~GeV in some regions on the parameter space for $c_L\sim c_R$. The impact of the chirality on the $13$~TeV reach can be studied in the same way, the result is shown on the right panel of Fig.~\ref{fig:2sslexclgen}. In this case we neglected the chirality dependence of the efficiencies and we included only the chirality effects on the single-production cross-section. The efficiencies are the same we used for
the purely right-handed coupling scenario discussed in Section~\ref{sec:efficiency}.

\section{A complete framework}\label{sec:completeframework}

In this section we extend the approach developed above to a general case with several light fermionic resonances. A scenario of this kind is very common in natural extensions of the SM, whose symmetry structure usually implies the presence of multiplets of light partners and not just single resonances. An example of such models are the minimal composite Higgs set-ups, which predict the existence of light top partners in complete $\SO(4)$ representations. For instance the $X_{5/3}$ resonance we considered in the previous section is usually one of the lightest states of an $\SO(4)$ quadruplet that includes two additional states with charge $2/3$, the $X_{2/3}$ and the $T$, and one state with charge $-1/3$, the $B$. The $X_{2/3}$ state is always nearly degenerate with the $X_{5/3}$, while the other two states are heavier, although the mass gap can be small~\cite{Panico:2011pw,DeSimone:2012fs} in some corners of the parameter space. Other exotic-charge partners could be considered, namely the $Y_{-4/3}$ and the $X_{8/3}$. The first one is usually a partner of the Bottom quark but it still couples to the Top even though, in most explicit models, with a reduced strength. The second originates from an enlarged Top Partner sector which contains an $\SO(4)$ $9$-plet \cite{Matsedonskyi:2014lla}.

The presence of several light states can be very useful to devise different complementary handles to probe the model. For this reason it is important to include all the relevant light states into the corresponding simplified description. In the following we show how this can be straightforwardly done in our framework at different levels of accuracy and, consequently, of complication. We only consider the case in which the resonances decay directly to SM states ignoring cascade decays, which is well justified by the following argument. Single Top Partner couplings to SM particles are always sizeable so that the direct decay to SM is always an allowed channel. Cascade decays can be relevant only in the presence of a considerable mass gap among the different Partners, otherwise they are suppressed or even forbidden by the small phase space. But if the gap is large the production cross section of the heavy state is much smaller than the one of the light resonance. The presence of the former can thus be safely ignored and the limit is driven by the lightest Partner decaying to SM particles. This rule would be violated if the sensitivity to the light resonance signal was much worse than the heavy one. We have not encountered a situation where this actually happens in the present context, nevertheless the addition of the couplings between the resonances in our simplified approach is straightforward and it could be easily implemented if needed.

Motivated by the minimal composite Higgs scenarios, we include in our effective description a set of resonances with electric charge $5/3$, $2/3$, $-1/3$ and $-4/3$, plus a model for the charge $8/3$ state borrowed from Ref.~\cite{Matsedonskyi:2014lla}. Leaving aside the charge $8/3$ partner, which we will not discuss any further referring the reader to Ref~\cite{Matsedonskyi:2014lla}, the relevant couplings are
\bea
\displaystyle
{\mathcal{L}}=&&\frac{g_w}2\left[c^{XV}_R \, \overline{X}_R\slashed{V}t_R+c^{XV}_L \, \overline{X}_L\slashed{V}t_L \right]+
\frac{g_w}2\left[c^{XV}_L \, \overline{X}_L\slashed{V}b_L + c^{XV}_R \, \overline{X}_R\slashed{V}b_R\right]  \nn \\
+&& \left[c^{Xh}_R \, h\,\overline{X}_L  t_R+c^{Xh}_L \, h\,\overline{X}_R  t_L\right]+ \left[c^{Xh}_L \, h\,\overline{X}_R b_L + c^{Xh}_R \, h\, \overline{X}_L  b_R \right] + \text{h.c.}
\,,
\label{eq:spc}
\eea
where $X$ generically denotes any of the top partners, $V=\{W^\pm,\,Z\}$ the EW gauge bosons and $h$ is the Higgs boson. Of course only the couplings respecting electric charge conservation are included. In the completely generic case, each resonance has an independent coupling to the SM particles, of arbitrary chirality. The strength of these interactions is parametrized, up to the $g_w/2$ normalization factor, by the dimensionless constants $c_{L/R}^{X\,V/h}$. For a single Top Partner all the phases can be reabsorbed by field redefinitions. Moreover interference effects
between different states are not relevant. Therefore the couplings can be assumed to be real in full generality.
In some models, additional derivative couplings involving the Higgs boson can also appear. However these interactions can be brought to a non-derivative form (at least at the trilinear level) by a field redefinition and incorporated in Eq.~(\ref{eq:spc}).
The Lagrangian in Eq.~(\ref{eq:spc}), plus of course the QCD interaction terms, is implemented in a {\sc{MadGraph}} model and is available at \cite{HEPMDB}.

\subsection{Production mechanisms}\label{sec:prod_mech}

\begin{figure}
\centering
\includegraphics[width=.675\textwidth]{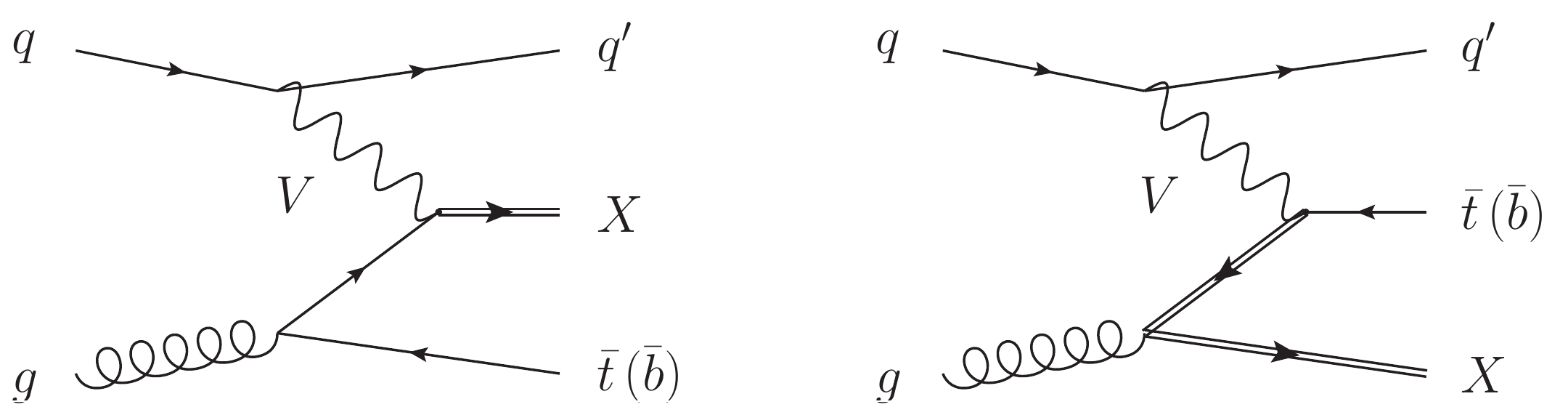}
\caption{The dominant diagrams contributing to the single production processes of a composite
resonance in association with a $t$ or $b$ quark.}
\label{fig:sp_gen}
\end{figure}

All the Partners can be pair-produced by QCD interactions. As we saw in the previous section, the corresponding cross sections are universal and can be parametrized by the $\sigma_{pair}(M_X)$ function which depends only on the resonance mass, $M_X$, reported in Table~\ref{tab:xsecpair}. The single production rate, on the other hand, depends not only on the partners masses, but also on their couplings to the SM quarks. Furthermore, two distinct single production processes can take place, we can either produce the Partner in association with a Top or with a Bottom quark. The corresponding tree-level diagrams are depicted in Figure~\ref{fig:sp_gen}. Notice that, due to the negligible coupling of the Higgs boson to the light SM quarks (including the Bottom), the interactions with the Higgs do not play a significant role in the production processes and are only relevant for the resonance decay.

\begin{table}[t]
\centering
\begin{tabular}{ c | c | c }
\rule[-6pt]{0pt}{1.75em} & \multicolumn{2}{|c}{$\sigma_{W^+ \overline{b}}+ \sigma_{W^- b}$ $[\textrm{fb}]$ @ NLO}\\
\hline
\rule[-6pt]{0pt}{1.75em}$M$\ [GeV] & $\sqrt{s}=8$ TeV & $\sqrt{s}=13$ TeV\\
\hline
\rule[-4pt]{0pt}{1.5em}600 &  (1490) 2100   & (6620) 9060  \\
\rule[-4pt]{0pt}{1.5em}700 &  (864) 1230   & (4240) 5820  \\
\rule[-4pt]{0pt}{1.5em}800 &  (514) 746   & (2810) 3860  \\
\rule[-4pt]{0pt}{1.5em}900 &  (317) 470   & (1910) 2720  \\
\rule[-4pt]{0pt}{1.5em}1000 &  (198) 298   & (1330) 1950  \\
\rule[-4pt]{0pt}{1.5em}1100 &  (127) 194   & (942) 1350  \\
\rule[-4pt]{0pt}{1.5em}1200 &  (82.1) 127 & (679) 982 \\
\rule[-4pt]{0pt}{1.5em}1300 &  (53.7) 84.8 & (493) 716 \\
\end{tabular}
\hspace{1cm}
\begin{tabular}{ c | c | c }
 \rule[-6pt]{0pt}{1.75em} & \multicolumn{2}{|c}{$\sigma_{W^+ \overline{b}} + \sigma_{W^- b}$ $[\textrm{fb}]$ @ NLO}\\
\hline
\rule[-6pt]{0pt}{1.75em}$M$\ [GeV] & $\sqrt{s}=8$ TeV & $\sqrt{s}=13$ TeV\\
\hline
\rule[-4pt]{0pt}{1.5em}1400 &  (35.5) 55.3   & (362) 540 \\
\rule[-4pt]{0pt}{1.5em}1500 &  (23.6) 37.5   & (268) 408 \\
\rule[-4pt]{0pt}{1.5em}1600 &  (15.9) 25.2   & (201) 305 \\
\rule[-4pt]{0pt}{1.5em}1700 & ---        & (151) 230  \\
\rule[-4pt]{0pt}{1.5em}1800 & ---        & (114) 174  \\
\rule[-4pt]{0pt}{1.5em}1900 & ---        & (87.4) 136  \\
\rule[-4pt]{0pt}{1.5em}2000 & ---        & (66.9) 102  \\
\multicolumn{3}{c}{\rule[-4pt]{0pt}{1.5em}}
\end{tabular}
\caption{NLO single production cross sections for the $Wb$ fusion for a unit coupling,
at $\sqrt{s}=8,13$ TeV (the LO values are in brackets), computed with MCFM \cite{mcfm} using the MSTW2008 parton distribution functions.}
\label{tab:xsWb}
\end{table}

As in the previous section, we parametrize the single production cross sections in a semi-analytic way as functions of the Top Partner couplings. The $t$-associated production is treated like in Eq.~(\ref{eq:prod1}) while the expression is simpler for the $b$-associated cross section. The latter is just proportional to $(c^{XV}_L)^2+(c^{XV}_R)^2$ since the interference term can be safely neglected due to the smallness of the Bottom quark mass. The cross-sections can be parametrized in full generality as
\bea
&&\sigma_{{\textrm{sing}}}(X\overline{t})=\left[\left(c^{XV}_L\right)^2+\left(c^{XV}_R\right)^2\right]\sigma_{V\overline{t}}(M_X) + c^{XV}_L\, c^{XV}_R \left( {m_t \over M_X + m_t} \right) \sigma^{\prime}_{V\overline{t}}(M_X) \,,\nonumber\\
&&\sigma_{\textrm{sing}}(X\overline{b})=\left[\left(c^{XV}_L\right)^2+\left(c^{XV}_R\right)^2\right] \sigma_{V\overline{b}}(M_X)\,,\nonumber\\
&&\sigma_{\textrm{sing}}(\overline{X}t)=\left[\left(c^{XV}_L\right)^2 +\left(c^{XV}_R\right)^2\right]{\sigma}_{\overline{V} t}(M_X)+ c^{XV}_L\, c^{XV}_R \left({m_t \over M_X + m_t} \right) {\sigma}^{\prime}_{\overline{V}t}(M_X) \,,\nonumber\\
&&\sigma_{\textrm{sing}}(\overline{X}b)=\left[\left(c^{XV}_L\right)^2+\left(c^{XV}_R\right)^2\right]{\sigma}_{\overline{V} b}(M_X)\,.
\label{eq:prod_gen_1}
\eea
in terms of the functions $\sigma_{V\bar{f}(\overline{V}f)}(M_X)$ and $\sigma_{V\bar{f}(\overline{V}f)}^{\prime}(M_X)$, with $f=t$ or $b$, which depend only on the resonance mass and not on the couplings. The labelling of $\sigma$ and $\sigma'$ reflects the fact that single production dominantly proceeds, as Figure~\ref{fig:sp_gen} shows, through the fusion of a vector boson $V$ with a gluon, producing the Partner and the associated $f$ or $\overline{f}$.\footnote{See also the discussion above Eq.~(\ref{eq:x_53_prod1}).}
Each function is easily computed, at the tree-level order, by a set of {\sc{MadGraph}} simulations. Some results are shown in Tables~\ref{tab:xsWt}, \ref{tab:xsWb} and \ref{tab:xsZt}, the sum of the Partner and anti-Partner rates are reported because the experimental searches typically collect positive and negative charge final states. No result is shown for $Z$-initiated processed producing a Bottom quark because single production vertexes with a $Z$ and a Bottom are typically suppressed in the Composite Higgs scenario. The interference functions are not reported in the Tables because it turns out that, as for the $X_{5/3}$ production discussed in Sec.~\ref{sec:x_53_chiral}, $\sigma'$ is well approximated (with a few percent error) by
\begin{equation}
\label{inta}
\sigma^{\prime}_{V\overline{t}(\overline{V}t)}(M_X) \simeq -5.2\, \sigma_{V\overline{t}(\overline{V}t)}(M_X)\,.
\end{equation}

A genuine NLO calculation of the single production processes is not currently available, however higher order QCD corrections might considerably affect the cross-section and we must find a way to estimate their impact. The MCFM code \cite{mcfm} is designed to compute QCD corrections to the SM single-Top production process, however it can also be used for BSM studies as it allows to change freely the mass of the Top and of the Bottom quark. By setting $m_t$ to $M_X$ we can  compute $\sigma_{W^+\overline{b}}$ and $\sigma_{W^-{b}}$, obtaining the results reported, together with the LO estimate within brackets, in Table~\ref{tab:xsWb}. The latter are almost exact NLO results, the only approximation being of having neglected Top loops, given that the SM Top plays now the role of the heavy partner. Similarly, by setting the Bottom mass to $M_X$ we computed $\sigma_{W^-\overline{t}}$ and $\sigma_{W^+t}$ in Table~\ref{tab:xsWb}. The other production modes initiated by a $W$ can not be obtained by MCFM, however they can be related to the previous ones by the following argument. The dominant Feynman amplitudes, depicted in Figure~\ref{fig:sp_gen}, are those with a $W$ in the $t$-channel emitted from the light quark line and interacting with the Top or with the Bottom producing the Partner. This structure is expected to be maintained at NLO because it physically reflects the fact that the process is approximately described by an on-shell $Wg$ fusion in accordance with the Effective $W$ approximation \cite{DeSimone:2012fs}. The amplitude thus factorizes in the $W$ emission term times the QCD matrix element of the single-production operator on an initial gluon and the final state $X$ $\overline{f}$ or $\overline{X}$ ${f}$. As far as QCD is concerned, all the Partners are identical and thus the matrix element is the same for all the partners with the same mass, what makes the difference for Partners of different charge is the $W$ emission which, after convoluting with the proton PDF's, is different for a $W^+$ and for a $W^-$. However QCD is also CP-invariant, which makes that the matrix element for $X$ $\overline{f}$ production operator is identical to the one for $\overline{X}$ $f$ production. This leads to the conclusion that the rates are identical, at least as far as the dominant diagrams are concerned, for the production of a Partner $X$ plus a SM anti-fermion $\overline{f}$ initiated by a $W^+$ and for the production of an anti-Partner $\overline{X}'$, of appropriate charge, plus the fermion $f$, again initiated by the $W^+$. The same obviously holds for the $W^-$. Therefore, we have
\begin{equation}
\displaystyle
\sigma_{W^\pm f}(M_X)\simeq\sigma_{W^\pm \bar{f}}(M_X)\,,
\end{equation}
and similarly for $\sigma'$. The above relations have been verified to hold at tree-level with good accuracy, by using them at NLO we finally obtain all the cross-sections for the $W$-initiated processes. In particular, we obtain $\sigma_{W^+ \overline{t}}$ and $\sigma_{W^- {t}}$ which are relevant for the $X_{5/3}$ and $\overline{X}_{5/3}$ production. By a similar argument we can also estimate the $Z$-initiated processes, which once again cannot be computed by MCFM. Because of the QCD symmetries the NLO corrections to the vector boson emission and the Top Partner production are expected not to depend on the vector boson charge, therefore to a good approximation the $K$-factor should be the same as for the $W$ initiated processes. The approximate NLO cross section in Table~\ref{tab:xsZt} are obtained by this assumption. For the interference term, which can not be estimated by MCFM where the coupling chirality is fixed, we rely on Eq.~(\ref{inta}).

Now that the cross-sections are known, the Top Partners decay Branching Ratios are the only missing theory inputs needed to compute the signal yield by the general formula in Eq.~(\ref{eq:nsignal}). The Branching Ratios are more complicated for a generic Top Partner than for the one of charge $5/3$ discussed in section~\ref{sec:x53}. In the latter case, $BR(X_{5/3}\to t W)=1$, while in general the Top Partner has several decay modes and the Branching Ratios carry a non trivial dependence on the Top Partner masses and couplings in Eq.~(\ref{eq:spc}). By computing the Branching ratios, which we report in Appendix~\ref{app:widths}, we complete our task of expressing the signal yield in an analytical form up to the experimental acceptance/efficiency factors $\epsilon_{n}$ in Eq.~(\ref{eq:nsignal}).

The general Top Partner model described above is rather complicated and it is characterized by a number of free parameters. It is thus worth stressing that our strategy does not require all the Partners being studied simultaneously and all the couplings being turned on and varied as free parameters. Different levels of complication are possible, depending on the accuracy one is aiming to reach and on extra physics assumptions one is willing to make. For instance, it is clear that each experimental search is going to be sensitive only to one or few Top Partner charges and production/decay topologies and only those should be considered for interpretation. On the other hand, the combination of different channels is possible, but not compulsory, with our approach. Moreover, not all the Top Partner couplings are expected to be equally sizable and furthermore rather generic correlations are expected among them. For example a charge-$2/3$ partner can couple and thus decay to $Wb$, $Zt$ and $Ht$ but the relative strength of the couplings, and thus the relative Branching Ratios, are not completely free parameters. In the case of a $\widetilde{T}$ singlet, the Branching Ratios are, respectively, $1/2$, $1/4$ and $1/4$ up to moderate model-dependent corrections. In the case of doublets, {\it{i.e.}} the $T$ or the $X_{2/3}$, the Branching Ratio to $Wb$ is suppressed and the other channels are approximately equal. One simplifying assumption could thus be to set the coupling ratio to these benchmark values and provide interpretation in the two hypotheses. Two examples of application of the general framework are discussed in the following section, the aim is to show how Top Partner search interpretation can be cast, at least to a first approximation, in simple $2$-dimensional coupling/mass plots analog to those for the $X_{5/3}$ in Figures~\ref{fig:2sslexcl_x53_R}, \ref{fig:2sslexclgen_13} and \ref{fig:2sslexclgen}.

\begin{table}[t]
\centering
\begin{tabular}{ c | c | c }
& \multicolumn{2}{|c}{$\sigma_{Z \overline{t}}+ \sigma_{Z t}$ $[\textrm{fb}]$ @ LO}\\
\hline
\rule[-6pt]{0pt}{1.75em}$M$\ [GeV] & $\sqrt{s}=8$ TeV & $\sqrt{s}=13$ TeV\\
\hline
\rule[-4pt]{0pt}{1.5em}600 &  (104) 128       & (588) 698  \\
\rule[-4pt]{0pt}{1.5em}700 &  (66.0) 82.8     & (411) 500  \\
\rule[-4pt]{0pt}{1.5em}800 &  (42.6) 54.7     & (295) 365  \\
\rule[-4pt]{0pt}{1.5em}900 &  (27.9) 36.7     & (214) 271  \\
\rule[-4pt]{0pt}{1.5em}1000 &  (18.7) 24.7    & (158) 203  \\
\rule[-4pt]{0pt}{1.5em}1100 &  (12.5) 16.9    & (118) 152  \\
\rule[-4pt]{0pt}{1.5em}1200 &  (8.45) 11.6  & (88.6) 116 \\
\rule[-4pt]{0pt}{1.5em}1300 &  (5.77) 8.00 & (67.4) 89.4 \\
\end{tabular}
\hspace{1cm}
\begin{tabular}{ c | c | c }
& \multicolumn{2}{|c}{$\sigma_{Z \overline{t}}+ \sigma_{Z t}$ $[\textrm{fb}]$ @ LO}\\
\hline
\rule[-6pt]{0pt}{1.75em}$M$\ [GeV] & $\sqrt{s}=8$ TeV & $\sqrt{s}=13$ TeV\\
\hline
\rule[-4pt]{0pt}{1.5em}1400 &  (3.95) 5.65     & (51.5) 69.2 \\
\rule[-4pt]{0pt}{1.5em}1500 &  (2.72) 3.94     & (39.6) 54.0 \\
\rule[-4pt]{0pt}{1.5em}1600 &  (1.87) 2.67     & (30.5) 42.0 \\
\rule[-4pt]{0pt}{1.5em}1700 & ---        & (23.7) 33.0  \\
\rule[-4pt]{0pt}{1.5em}1800 & ---        & (18.5) 25.9  \\
\rule[-4pt]{0pt}{1.5em}1900 & ---        & (14.5) 20.5  \\
\rule[-4pt]{0pt}{1.5em}2000 & ---        & (11.4) 16.2  \\
\multicolumn{3}{c}{\rule[-4pt]{0pt}{1.5em}}
\end{tabular}
\caption{Single production cross sections for the $Z t$ fusion for a unit coupling,
at $\sqrt{s}=8,13$ TeV. The LO values (in brackets) have been computed with {\sc MadGraph}
using the MSTW2008 parton distribution functions.
The NLO values are obtained by multiplying LO by the $k$-factors obtained
for $W t$ fusion (see Table~\ref{tab:xsWt}).}
\label{tab:xsZt}
\end{table}

\subsection{Applications}

In this subsection we present two simple applications of the general framework.
 In the first example we reinterpret the current searches for
charge-$2/3$ resonances. Afterwards we discuss how in our formalism one can easily handle
a typical scenario in which two resonances contribute to the same final state.
These two examples are motivated by the usual Composite Higgs scenarios. Indeed, in minimal
models of this kind, the lightest top partner can be either an exotic state with charge $5/3$ that
is part of an $\SO(4)$ quadruplet, or a charge $2/3$ state which is an $\SO(4)$ singlet.
The analyses presented in the following are thus typically the ones leading to the most
constraining bounds on the composite Higgs parameter space.

\subsubsection[The $\widetilde{T}$ singlet]{The $\mathbf{\widetilde{T}}$ singlet}\label{sec:charge23}

The $\widetilde{T}$ singlet is easily described within our framework. It is a charge $2/3$ Partner, denoted as ``$T$'' in our model, characterized by a sizable $c^{TW}$ coupling with the Bottom and thus copiously produced in association with a Bottom quark. Single production with a Top is also possible, but relatively suppressed by the larger Top quark mass and thus in many cases negligible. The coupling is Left--Handed to a very good approximation because the Right--Handed Bottom has a small compositeness fraction and thus feeble interactions with the Partners. It also couples to $Zt$ and $ht$ with considerable strength and thus it decays to $Wb$, $Zt$ or $ht$. Describing the $\widetilde{T}$ phenomenology in full generality thus requires a number of free parameters, namely $5$ couplings plus the mass. While this is straightforward and technically doable in our framework, a simpler treatment is possible. Indeed, out of these $5$ couplings only $3$ combinations matter, namely the single production coupling $c^{TW}_L$ and the two Branching Ratios $BR(ht)$ and $BR(Zt)$ which only depend on the $c^{TZ}$ and $c^{Th}$ overall strength and not on their chiralities.\footnote{We are ignoring here the possible dependence of the acceptance on the coupling chirality.} An even simpler but still accurate enough approach, which we adopt in what follows, is to ignore the coupling dependence of the Branching Ratios and to set them to the ``typical'' values for a SM singlet, namely $BR(T \rightarrow Wb) = 1/2$ and $BR(T \rightarrow Zt) = BR(T \rightarrow ht) = 1/4$. In most models this approximation is accurate to $10\%$ level \cite{DeSimone:2012fs} and considerable departures might occur only in corners of the parameter space. By this assumption, the relevant parameter space is reduced to the two--dimensional plane $(M_T,c^{TW}_L)$.

\begin{figure}
\centering
\includegraphics[width=0.475\textwidth]{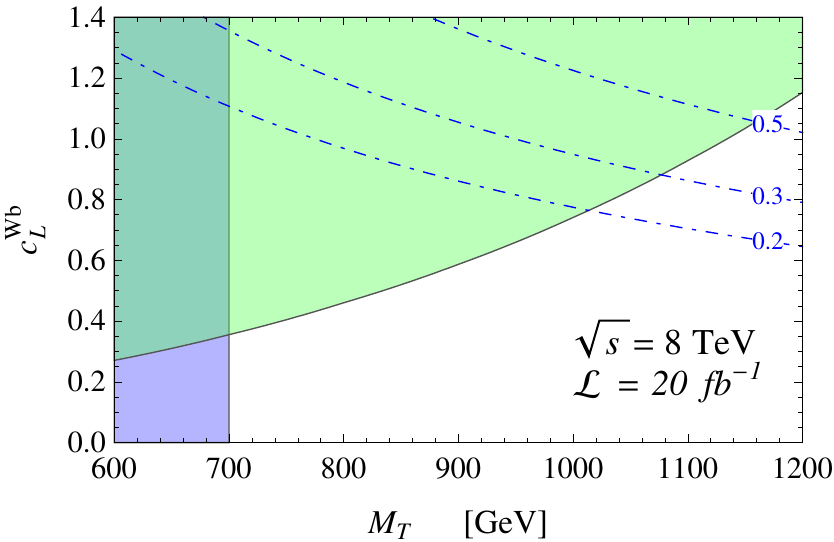}
\hfill
\includegraphics[width=0.475\textwidth]{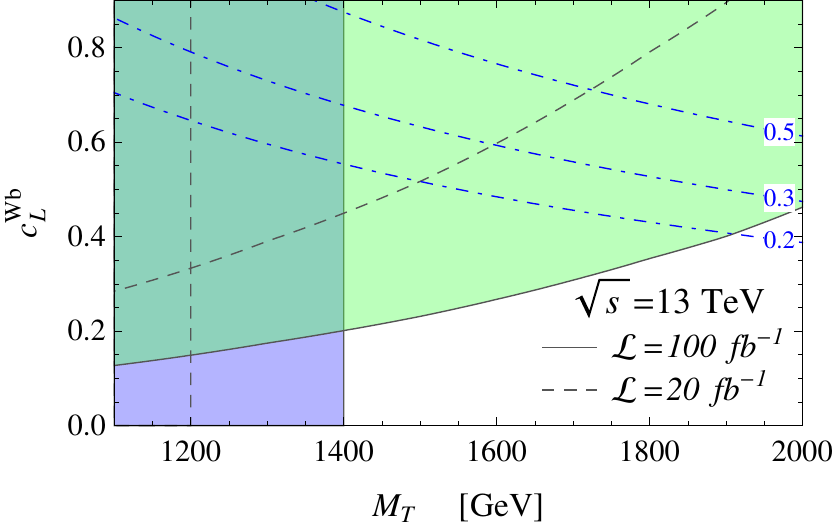}
\caption{Current bounds (left panel)
on the mass of a charge-$2/3$ state decaying with $50\%$ branching ratio into $W b$.
The bounds are presented for different values
of the coupling $c_L$ to the bottom quark.
The gray shaded area is excluded from
pair production only,
the green shaded area corresponds to the estimated exclusion from $b$-associated single
production~\cite{Ortiz:2014iza}.
In the right panel: estimated projection of the bounds for the $13\ \mathrm{TeV}$ LHC run.
The dash-dotted blue lines show the contours with $\Gamma/M = 0.2, 0.3, 0.5$.}
\label{fig:excl_tb}
\end{figure}

So far the strongest bounds presented by the experimental collaborations
were derived in the CMS analysis in Ref.~\cite{Chatrchyan:2013uxa}, which considers a generic
charge-$2/3$ resonance decaying into $Wb$, $Zt$ and $ht$.\footnote{Other experimental
searches for charge-$2/3$ resonances performed by the ATLAS
collaboration are available in the literature. In particular searches for resonances decaying
into a single channel ($Wb$~\cite{TheATLAScollaboration:2013sha}, $Zt$~\cite{TheATLAScollaboration:2013oha} and $ht$~\cite{ATLAS:2013ima}) have been presented,
as well as searches for resonances giving rise to final states with two same-sign
leptons~\cite{atlas051}.
The bounds obtained in all these studies, however, are weaker than the ones
of Ref.~\cite{Chatrchyan:2013uxa}, thus we will only use the latter for our analysis.}
The bounds are based on pair
production only and are presented as a function of the branching ratios into the three
decay channels. Depending on the branching ratios, the
lower bound on the mass of the resonance ranges from $687\ \mathrm{GeV}$ to
$782\ \mathrm{GeV}$. For the configuration we consider ($BR(T \rightarrow Wb) = 1/2$) the bound
is $M_T \gtrsim 700\ \mathrm{GeV}$.
Although single production has not been included in the experimental analyses so far,
its cross section can be sizable and can easily become larger than the pair production one,
especially for large resonance masses. Unfortunately the present CMS and ATLAS analyses seem to be
targeted exclusively on pair production, in such a way that a recast to include single production
is not doable. To get an idea of how much the single production process can improve the pair production
bounds we thus focus on the analysis of Ref.~\cite{Ortiz:2014iza} and reinterpret their results.
For our reinterpretation we extracted from the results of Ref.~\cite{Ortiz:2014iza} the number of signal events
needed for the exclusion ($S_{exc} = 26$) and the cut efficiency.
Unfortunately the data included in Ref.~\cite{Ortiz:2014iza} allows us to extract the cut efficiency
only for one mass point, thus in our reinterpretation we assume that it is roughly independent of the resonance mass.
The results of our analysis are shown in Fig.~\ref{fig:excl_tb}.
The plots show that, in the case of the $8\ \mathrm{TeV}$ LHC searches,
for small values of the single production coupling ($c_L \lesssim 0.3$)
the strongest bounds
come from pair production. For larger values, instead, single production leads to a bound that
steeply increases with $c_L$ and reaches $M_T \gtrsim 1\ \mathrm{TeV}$ for $c_L \simeq 0.7$.
To obtain the projections for the $13\ \mathrm{TeV}$ LHC run, we assume that
the number of events needed for the exclusion and the cut efficiencies coincide with the
$8\ \mathrm{TeV}$ ones. The result is shown in the right panel of Fig.~\ref{fig:excl_tb}.

\subsubsection{A two-Partners interpretation}
\label{2P}

As a final example in this subsection we consider one scenario in which two resonances
can contribute to the same final state. This possibility is not uncommon in explicit models
in particular in the composite Higgs framework.
A typical example, on which we will focus in the following, is the case in which a charge 5/3 state ($X_{5/3}$)
is present together with a charge $-1/3$ resonance ($B$).
Both resonances contribute to final states with two same-sign leptons, moreover the signal
efficiencies for the two states are similar.\footnote{This was verified for
$7\ \mathrm{TeV}$ collider energy in Ref.~\cite{DeSimone:2012fs}.}
For our illustrative purposes it is thus reasonable to simplify the analysis
by assuming the same cuts acceptances for both states.
A more rigorous study, of course, will require a separate determination of the $B$ state
acceptances. Some difference with respect to the $X_{5/3}$ events can be expected, for instance,
in the lepton distributions in single production.

The number of signal events can be easily computed from Eq.~(\ref{eq:nsignal})
by summing over the various production channels of the two resonances:
\bea
N_{signal} &=& {\cal L} \Big[
\textrm{BR}_{s.p.} \, \epsilon_{s.p.} (M_X) \, \sigma_{s.p.}(M_X) + \textrm{BR}_{p.p.} \, \epsilon_{p.p.} (M_X) \, \sigma_{p.p.}(M_X) \nn \\
&& +\, \textrm{BR}_{s.p.} \, \epsilon_{s.p.} (M_B) \, \sigma_{s.p.}(M_B) + \textrm{BR}_{p.p.} \, \epsilon_{p.p.} (M_B) \, \sigma_{p.p.}(M_B) \Big]\,.
\label{eq:nsignal_XB}
\eea
In order to simplify the analysis, we will assume a specific pattern for the resonances couplings motivated by the minimal
composite Higgs scenarios.\footnote{A detailed discussion on this point can be found in Ref.~\cite{Matsedonskyi_inpreparation}.}
Although the $B$ is in principle allowed to decay in three different channels ($Wt$, $Zb$ and $Hb$),
we will assume that the $W t$ decay mode dominates over the rest and take $BR(B\to Wt) = 1$.
Moreover we will assume that the $X_{5/3}$ and $B$ resonances are coupled to the $t_R$ quark only and the
corresponding coupling strengths are equal: $c_R^{BW}=c_R^{XW}$.
With these choices we are left with just three free parameters, namely the mass of the $X_{5/3}$ state $M_X$,
the mass gap between the two resonance $\Delta \equiv M_B - M_X > 0$, which we assume to be positive,
and one coupling $c_R \equiv c_R^{XW} = c_R^{BW}$.

\begin{figure}
\centering
\includegraphics[width=0.47\textwidth]{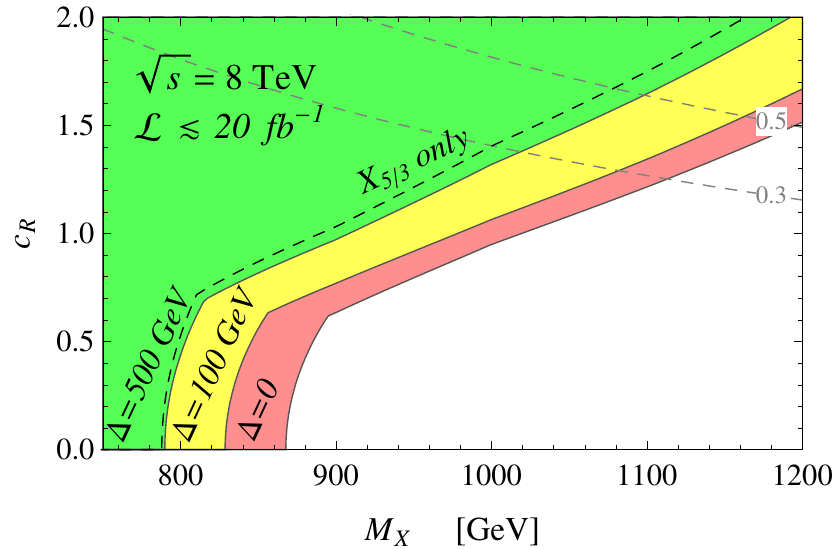}
\hfill
\includegraphics[width=0.47\textwidth]{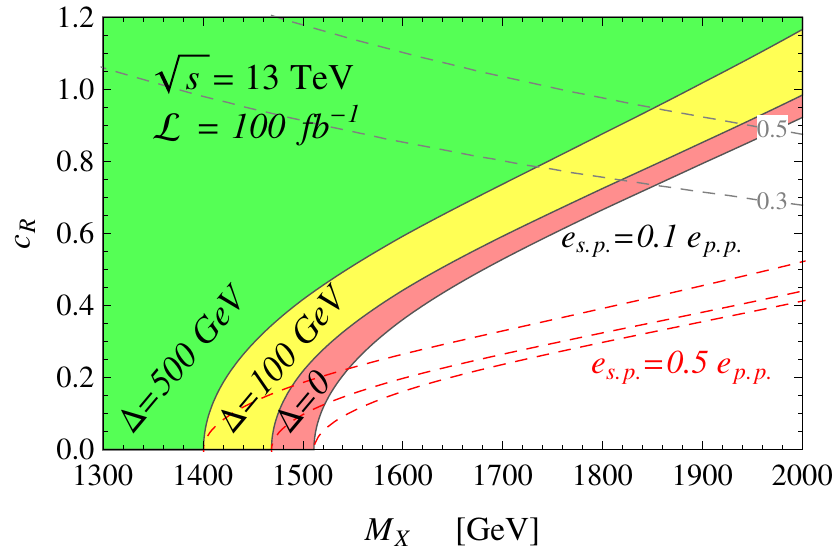}
\caption{Bounds on the mass and couplings of a charge-$5/3$ state
in a presence of an additional resonance ($B$) contributing to the same final states.
The excluded regions (red, yellow and green areas) correspond to a mass split $\Delta = 0, 100, 500\ \mathrm{GeV}$.
Left panel: bounds obtained by using the $\sqrt s = 8$~TeV data (the dashed black line corresponds
to the limit with only the $X_{5/3}$ resonance). Right panel: expected exclusion for $\sqrt s = 13$~TeV
and integrated luminosity $100$~fb$^{-1}$, assuming $e_{s.p.} = 0.1\, e_{p.p.}$ (colored regions)
and $e_{s.p.} = 0.5\, e_{p.p.}$ (red dashed lines).
The dashed gray lines show the contours with $\Gamma(X_{5/3})/M_X = 0.3, 0.5$.}
\label{fig:excl_XB}
\end{figure}

In Fig.~\ref{fig:excl_XB} we show the current bounds and the expected future LHC reach on the parameter space of our simplified model.
One can see that if the $B$ is $500\ \mathrm{GeV}$ heavier than the $X_{5/3}$ its contribution to the signal cross section is
almost negligible and we basically recover the result shown in Fig.~\ref{fig:2sslexcl_x53_R}. When the resonances are exactly
degenerate, instead, the signal cross section is doubled, leading to an enhancement of the bounds of order $100\ \mathrm{GeV}$.
A mild mass gap (of order of $100\ \mathrm{GeV}$) is already enough to suppress significantly the role of the
$B$ state. In this case the increase in the bounds is of order $50\ \mathrm{GeV}$, that is around one half of the
increase we found in the degenerate case.


\section{Prospects at a future 100 TeV collider}\label{sec:100TeV}

As a last topic, in this section we provide a rough analysis of the reach of a hypothetical
$100$~TeV hadronic collider. For definiteness we focus on two benchmark scenarios. The first one is the
set-up in section~\ref{sec:x53_simplest} containing only an exotic charge-$5/3$
resonance that couples dominantly with the $t_R$ field. The second scenario is the one
we discussed in Section~\ref{sec:charge23} with only
a charge-$2/3$ resonance with $50\%$ branching ratio into $Wb$.

\begin{table}[t]
\begin{center}
\begin{tabular}{c|c}
& \multicolumn{1}{c}{$\sigma_{pair}$\ [fb] @ LO}\\
\hline
\rule[-6pt]{0pt}{1.75em}$M$\ [TeV] & $\sqrt{s} = 100$ TeV \\
\hline
\rule[-4pt]{0pt}{1.5em}2& 520\\
\rule[-4pt]{0pt}{1.5em}3& 62.0 \\
\rule[-4pt]{0pt}{1.5em}4& 12.7\\
\rule[-4pt]{0pt}{1.5em}5& 3.49\\
\rule[-4pt]{0pt}{1.5em}6& 1.15\\
\rule[-4pt]{0pt}{1.5em}7& 0.430\\
\rule[-4pt]{0pt}{1.5em}8& 0.175
\end{tabular}
\hspace{.5cm}
\begin{tabular}{c|c}
& \multicolumn{1}{c}{$\sigma_{pair}$\ [fb] @ LO}\\
\hline
\rule[-6pt]{0pt}{1.75em}$M$\ [TeV] & $\sqrt{s} = 100$ TeV \\
\hline
\rule[-4pt]{0pt}{1.5em}9& 0.0761\\
\rule[-4pt]{0pt}{1.5em}10& 0.0346\\
\rule[-4pt]{0pt}{1.5em}11& 0.0164\\
\rule[-4pt]{0pt}{1.5em}12& 0.00796\\
\rule[-4pt]{0pt}{1.5em}13& 0.00393\\
\rule[-4pt]{0pt}{1.5em}14& 0.00198\\
\rule[-4pt]{0pt}{1.5em}15& 0.00101
\end{tabular}
\hspace{.5cm}
\begin{tabular}{c|c}
& \multicolumn{1}{c}{$\sigma_{pair}$\ [fb] @ LO}\\
\hline
\rule[-6pt]{0pt}{1.75em}$M$\ [TeV] & $\sqrt{s} = 100$ TeV \\
\hline
\rule[-4pt]{0pt}{1.5em}16& 0.000513\\
\rule[-4pt]{0pt}{1.5em}17& 0.000263\\
\rule[-4pt]{0pt}{1.5em}18& 0.000135\\
\rule[-4pt]{0pt}{1.5em}19& 0.0000695\\
\rule[-4pt]{0pt}{1.5em}20& 0.0000356\\
\rule[-4pt]{0pt}{1.5em}21& 0.0000181\\
\multicolumn{2}{c}{\rule[-4pt]{0pt}{1.5em}}
\end{tabular}
\caption{Top partners pair production cross
section (in fb), for $\sqrt{s}=100$ TeV, computed at LO with \textsc{MadGraph},
using the cteq6 parton distribution functions.}
\label{tab:xsecpair100}
\end{center}
\end{table}

\begin{table}[t]
\begin{center}
\begin{tabular}{c|c}
& \multicolumn{1}{c}{$\sigma_{W^+ \overline{t}} + \sigma_{W^- t}$\ [pb] @ LO}\\
\hline
\rule[-6pt]{0pt}{1.75em}$M$\ [TeV] & $\sqrt{s} = 100$ TeV \\
\hline
\rule[-4pt]{0pt}{1.5em}2& 19.6\\
\rule[-4pt]{0pt}{1.5em}3& 8.84\\
\rule[-4pt]{0pt}{1.5em}4& 4.62\\
\rule[-4pt]{0pt}{1.5em}5& 2.62\\
\rule[-4pt]{0pt}{1.5em}6& 1.57\\
\rule[-4pt]{0pt}{1.5em}7& 0.985\\
\rule[-4pt]{0pt}{1.5em}8& 0.642\\
\rule[-4pt]{0pt}{1.5em}9& 0.427\\
\rule[-4pt]{0pt}{1.5em}10& 0.287\\
\rule[-4pt]{0pt}{1.5em}11& 0.197
\end{tabular}
\hspace{1.cm}
\begin{tabular}{c|c}
& \multicolumn{1}{c}{$\sigma_{W^+ \overline{t}} + \sigma_{W^- t}$\ [pb] @ LO}\\
\hline
\rule[-6pt]{0pt}{1.75em}$M$\ [TeV] & $\sqrt{s} = 100$ TeV \\
\hline
\rule[-4pt]{0pt}{1.5em}12& 0.138\\
\rule[-4pt]{0pt}{1.5em}13& 0.0970\\
\rule[-4pt]{0pt}{1.5em}14& 0.0695\\
\rule[-4pt]{0pt}{1.5em}15& 0.0499\\
\rule[-4pt]{0pt}{1.5em}16& 0.0360\\
\rule[-4pt]{0pt}{1.5em}17& 0.0261\\
\rule[-4pt]{0pt}{1.5em}18& 0.0191\\
\rule[-4pt]{0pt}{1.5em}19& 0.0141\\
\rule[-4pt]{0pt}{1.5em}20& 0.0104\\
\rule[-4pt]{0pt}{1.5em}21& 0.00767
\end{tabular}
\caption{Single production cross sections (in pb) for the Wt fusion channel for a unit coupling,
for $\sqrt{s}=100$ TeV, computed at LO with \textsc{MadGraph},
using the cteq6 parton distribution functions.}
\label{tab:xsecsingleWt100}
\end{center}
\end{table}

\begin{table}[t]
\begin{center}
\begin{tabular}{c|c}
& \multicolumn{1}{c}{$\sigma_{W^+ \overline{b}} + \sigma_{W^- b}$\ [pb] @ LO}\\
\hline
\rule[-6pt]{0pt}{1.75em}$M$\ [TeV] & $\sqrt{s} = 100$ TeV \\
\hline
\rule[-4pt]{0pt}{1.5em}2& 76.3\\
\rule[-4pt]{0pt}{1.5em}3& 32.5\\
\rule[-4pt]{0pt}{1.5em}4& 16.4\\
\rule[-4pt]{0pt}{1.5em}5& 9.12\\
\rule[-4pt]{0pt}{1.5em}6& 5.43\\
\rule[-4pt]{0pt}{1.5em}7& 3.38\\
\rule[-4pt]{0pt}{1.5em}8& 2.17\\
\rule[-4pt]{0pt}{1.5em}9& 1.44\\
\rule[-4pt]{0pt}{1.5em}10& 0.970\\
\rule[-4pt]{0pt}{1.5em}11& 0.668
\end{tabular}
\hspace{1.cm}
\begin{tabular}{c|c}
& \multicolumn{1}{c}{$\sigma_{W^+ \overline{b}} + \sigma_{W^- b}$\ [pb] @ LO}\\
\hline
\rule[-6pt]{0pt}{1.75em}$M$\ [TeV] & $\sqrt{s} = 100$ TeV \\
\hline
\rule[-4pt]{0pt}{1.5em}12& 0.466\\
\rule[-4pt]{0pt}{1.5em}13& 0.327\\
\rule[-4pt]{0pt}{1.5em}14& 0.233\\
\rule[-4pt]{0pt}{1.5em}15& 0.167\\
\rule[-4pt]{0pt}{1.5em}16& 0.121\\
\rule[-4pt]{0pt}{1.5em}17& 0.0881\\
\rule[-4pt]{0pt}{1.5em}18& 0.0642\\
\rule[-4pt]{0pt}{1.5em}19& 0.0472\\
\rule[-4pt]{0pt}{1.5em}20& 0.0348\\
\rule[-4pt]{0pt}{1.5em}21& 0.0257
\end{tabular}
\caption{Single production cross sections (in pb) for the Wb fusion channel for a unit coupling,
for $\sqrt{s}=100$ TeV, computed at LO with \textsc{MadGraph},
using the cteq6 parton distribution functions.}
\label{tab:xsecsingleWb100}
\end{center}
\end{table}

\begin{figure}[t]
\centering
\includegraphics[width=0.48\textwidth]{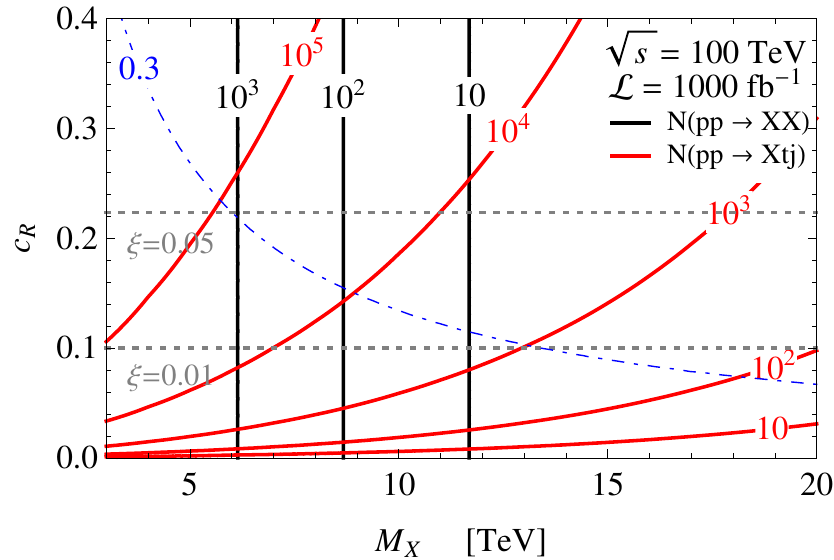}
\hfill
\includegraphics[width=0.48\textwidth]{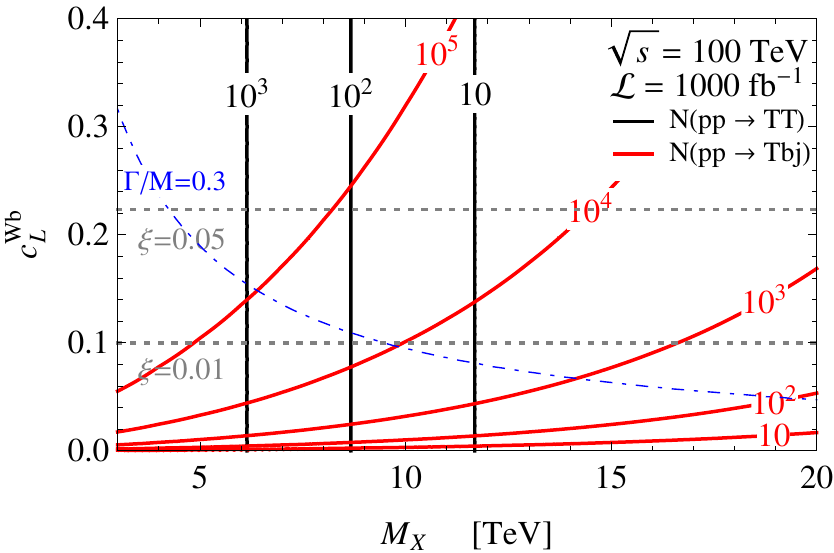}
\caption{Number of events for pair and single production of a charge-$5/3$ state (left panel)
and a charge-$2/3$ state coupled to the $b$ quark (right panel) as
a function of the single production couplings. To obtain the results we assumed
$\sqrt{s} = 100\ \mathrm{TeV}$ collider energy and
$L = 1000\ \mathrm{fb}^{-1}$ integrated luminosity.
The dotted gray lines show the typical size of the single production coupling
for $\xi = 0.05$ and $\xi = 0.01$.
The dash-dotted blue line denotes the contour with $\Gamma/M = 0.3$.}
\label{fig:x53_events_100}
\end{figure}

The production cross sections for pair production and for single production (in association
with a $t$) are listed in Tables~\ref{tab:xsecpair100}, \ref{tab:xsecsingleWt100} and \ref{tab:xsecsingleWb100}.
The results have been computed at LO with \textsc{MadGraph} by using the cteq6 parton
distribution functions. In Fig.~\ref{fig:x53_events_100} we show the number
of events in the two production channels as a function of the mass of the resonance and of the
single production coupling for $L = 1\ \mathrm{ab}^{-1}$ integrated luminosity. As can be seen from the plot,
pair production becomes essentially irrelevant above $m_X \simeq 10\ \mathrm{TeV}$. To access particles
masses above this scale one must therefore rely on single-production processes.

\begin{figure}[t]
\centering
\includegraphics[width=0.48\textwidth]{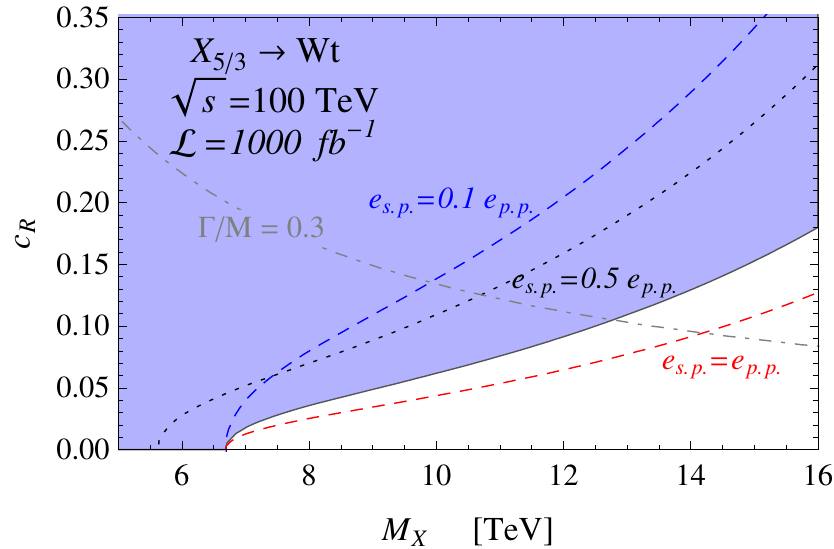}
\hfill
\includegraphics[width=0.48\textwidth]{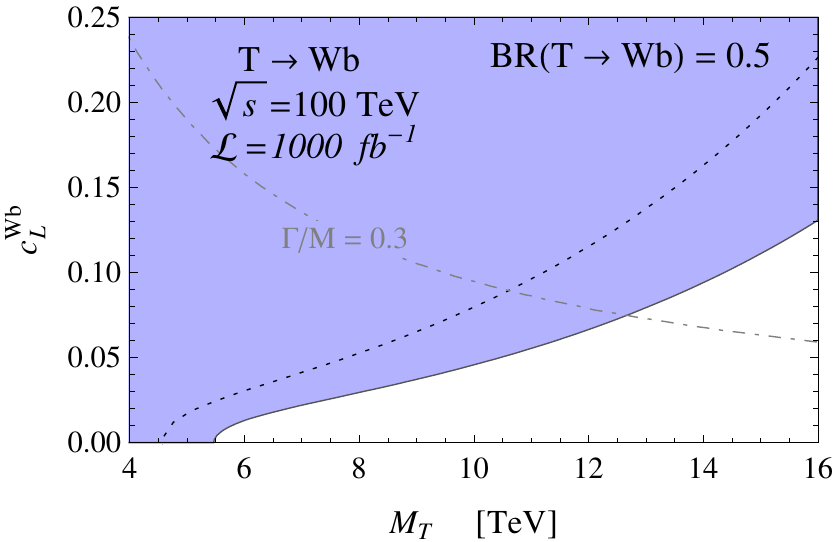}
\caption{Estimated exclusion bounds on the mass of a charge-$5/3$ state decaying exclusively
to $W t$ (left panel) and of a charge-$2/3$ state decaying into $Wb$ with $50\%$ branching ratio.
To obtain the excluded regions we assumed $\sqrt{s} = 100\ \mathrm{TeV}$ collider energy and
$L = 1000\ \mathrm{fb}^{-1}$ integrated luminosity. For the $X_{5/3}$ exclusions (left panel)
the solid and dashed curves are obtained by assuming
$S_{exc} = 10$ for different values of the single production efficiency $e_{s.p.} = 0.1\, e_{p.p.}$ (blue curve),
$e_{s.p.} = 0.5\, e_{p.p.}$ (black curve) and $e_{s.p.} = e_{p.p.}$ (red curve).
The dotted black line corresponds to $S_{exc} = 30$ and $e_{s.p.} = 0.5\, e_{p.p.}$.
For the charge-$2/3$ resonance exclusion (right panel) we assumed the same efficiency for single and pair production
($e_{p.p.} = e_{s.p.} = 0.012$) and $S_{exc} = 25$ (solid curve) and $S_{exc} = 75$ (dashed curve).
In both plots the dash-dotted gray line shows the contour with $\Gamma/M = 0.3$.}
\label{fig:2sslexclgen_100}
\end{figure}

To get a rough idea of the reach of the $100\ \mathrm{TeV}$ machine, we repeat the analyses that we performed
in the previous sections.
For the $X_{5/3}$ case we can focus
on the $2ssl$ channel and use a simple naive recast of the preliminary $14\ \mathrm{TeV}$ analysis of Ref.~\cite{cms14}.
We assume that the efficiency for extracting the signal in pair production channels is the same, namely $e_{p.p} = 0.017$,
and does not depend on the resonance mass. Moreover we assume that the number of signal events needed for the
exclusion is roughly unchanged, $S_{exc} \simeq 10$.\footnote{Notice that the efficiency $e_{p.p.}$
and the number of signal events we use for exclusion $N_{bound}$
are also close to the ones for the $8\ \mathrm{TeV}$ LHC (see section~\ref{sec:efficiency}).}
For single production we focus on the three
benchmark scenarios with $e_{s.p.} = 0.1\, e_{p.p.}$, $e_{s.p.} = 0.5\, e_{p.p.}$ and $e_{s.p.} = e_{p.p.}$.
The estimate for the bounds are shown in the left panel of Fig.~\ref{fig:2sslexclgen_100} for an integrated
luminosity $L = 1\ \mathrm{ab}^{-1}$. In the plot we also show how the bound changes in the
more pessimistic scenario with $S_{exc} = 30$ (dotted black line).

For the case of a charge-$2/3$ resonance we consider the procedure used in Section~\ref{sec:charge23}.
We assume that the number of signal events needed for the exclusion is roughly equal to the
ones needed at $8\ \mathrm{TeV}$ ($S_{exc} \simeq 25$) and that the efficiency is the same
for pair and single production ($e_{p.p} = e_{s.p.} = 0.012$).
The estimate for the bounds are shown in the right panel of Fig.~\ref{fig:2sslexclgen_100}.
In the plot we also show how the bound changes in the
more pessimistic scenario with $S_{exc} = 75$ (dotted black line).

The estimated bound on the charge-$5/3$ and charge-$2/3$ resonances are roughly comparable.
In the case of purely pair production resonance masses around $M \simeq 6\ \mathrm{TeV}$ can be tested.
If single production becomes sizable (for $c \gtrsim 0.1$) the bounds can easily reach $M \gtrsim 12\ \mathrm{TeV}$.
Notice that testing resonances with larger masses through single production can become increasingly
difficult above $M \simeq 12\ \mathrm{TeV}$ because the largish single production couplings needed for a
sizable cross section also imply a large resonance width.

To conclude the discussion we comment on the typical size of the single production couplings
that will be plausible to consider at a $100\ \mathrm{TeV}$ collider. After the full LHC program
we will be presumably able to test values of Higgs compositeness of the order $\xi = (v/f)^2 \gtrsim 0.1$,
both through single Higgs production measurements and direct resonances searches. Unfortunately any
hadronic machine can not significantly improve the precision on single Higgs processes, thus,
in the absence of future leptonic colliders the bound on $v/f$ will remain presumably similar
also at a $100\ \mathrm{TeV}$ collider. In this situation a value $\xi = 0.05$
(corresponding to couplings  $c \sim v/f = 0.22$\footnote{See Section~\ref{sec:conclusions} for
more details about the couplings estimate.}) could be considered as a reasonable benchmark point.
On the other hand, if future leptonic collider experiments will be performed, the precision
on single Higgs measurements can drastically increase and values of Higgs compositeness
$\xi \sim 0.005$ could be testable (see for instance~\cite{futureprecision}).
In this scenario a benchmark point $\xi = 0.01$ (corresponding to couplings $c \sim v/f = 0.1$) could be
realistic.


\section{Conclusions and Outlook}\label{sec:conclusions}

In this paper we described a strategy for the interpretation of Top Partners collider searches addressing the model-dependence issue which characterizes the phenomenology of this kind of resonances. We hope that it could guide the experimental collaborations in the preparation of Run-$2$ LHC searches. Our philosophy is conveniently summarized by comparing it with an alternative approach developed in Ref.~\cite{Barducci:2014ila} and implemented in the computer package XQCAT. The latter consists of an automated recasting tool which incorporates publicly available experimental data and reinterprets them within general Top Partner models. Our strategy is basically opposite to the one of Ref.~\cite{Barducci:2014ila}, we have designed it to avoid recasting, allowing the experimental collaborations to carry on the data interpretation autonomously by setting limits on a Simplified Model parameter space. The Simplified Model limits are easy to interpret within concrete models, in a way that requires no recasting and no knowledge of the experimental details of the analyses. Furthermore, in the fortunate case of a discovery the usage of a Simplified Model will become an unavoidable intermediate step to characterize the excess, also by comparing different channels, towards the identification of the ``true'' microscopic theory. Though based on the opposite philosophy, the approach of Ref.~\cite{Barducci:2014ila} is complementary to ours. Indeed by Simplified Models we can cover most of the relevant Physics scenarios involving Top Partners and the approach could be extended (see below) to other interesting particles, but we will definitely be unable to cover the most exotic models, including those that might emerge by future theoretical speculations. For the latter, recasting might eventually be needed. Notice also that our limit-setting strategy facilitates recasting, especially if the experimental collaborations will also report the intermediate steps, namely the efficiencies for the individual signal topologies. The latter could be useful also in other contexts which are not directly described by our Simplified Model.

The Simplified Model is defined by Eq.~(\ref{eq:spc}), which can be used to describe different Top Partner species and different signal topologies. The theoretical tools which are needed to study the model, namely the production rates and the Branching Ratios, are reported in Section~\ref{sec:prod_mech} and in Appendix~\ref{app:widths}. A {\sc MadGraph} implementation of the model, designed to simulate the Top Partners signals and to extract the efficiencies, is briefly described in Appendix~\ref{app:Madgraph_model} and publicly available. As concrete applications of the method, we studied $X_{5/3}$ and ${\widetilde{T}}$ single and pair production, we also studied the combined effects of $B$ and $X_{5/3}$ Partners in \emph{2ssl} final states. In each case we performed a theory recasting of the available $8$~TeV Run-$1$ results and an estimate of the $13$~TeV Run-$2$ reach. We showed how the results, reported in Figures~\ref{fig:2sslexcl_x53_R}, \ref{fig:2sslexclgen_13} and \ref{fig:excl_tb}, can be conveniently expressed in a simple mass--coupling plane under minor and well-justified theoretical assumptions. We also showed, in the case of the $X_{5/3}$ Partner, how easily one can go beyond the two-parameter interpretation by including the effect of the single production coupling chirality on the production rate and on the efficiencies. The result is summarized in Figure~\ref{fig:2sslexclgen}. Finally, a rough estimate of the reach at a hypothetical $100$~TeV collider is performed in Section~\ref{sec:100TeV}.

On top of serving as an illustration of the interpretation strategy, our result also provides an assessment of the current Top Partner limits and of the future prospects. In order to evaluate them quantitatively, in terms of a mass reach, we need an estimate of the $c^{XV}$ couplings to vector bosons which control the single production rate. The size of the latter couplings can vary considerably in different models, and even in the context of the CH scenario their parametric scaling is not fixed, it depends on the Top Partner species and on the detailed implementation of Partial Compositeness in the Top sector. A detailed estimate, and a quantitative assessment of the limits in explicit CH models will be presented in Ref.~\cite{Matsedonskyi_inpreparation}. However, a simple generic estimate goes as follows. The single production couplings are necessarily proportional to the EWSB scale $v$ because the gauge interactions are flavor diagonal if the EW symmetry is unbroken. In CH any $v$ insertion is accompanied by $1/f$, where $f$ is the Goldstone boson Higgs $\sigma$-model scale, therefore the couplings are proportional to the universal factor
\begin{equation}
c^{XV}_{L,R} \propto \frac{v}{f} = \sqrt{\xi}\,.\nonumber
\end{equation}
Given that $\xi\sim 0.1$ in reasonably natural and viable CH models, the above estimate suggests a typical value of $0.3$ for the single production couplings even though considerable numerical enhancements are possible in explicit models. For such a value, our results show that single production has a marginal impact on the $8$~TeV Top Partners mass limit but it becomes important for the Run-$2$ reach. It must also be noticed that our estimate of the single-production reach is most likely a conservative one because it is not based on sound and well optimized experimental studies. We believe that the actual Run-$2$ searches might achieve a better sensitivity.

The present work could be extended in the following directions. First of all, other Top Partners might be searched for, in the same final states discussed in this paper or in other ones. We focused on $X_{5/3}$ and ${\widetilde{T}}$, which as of now we regard as the most promising signatures of CH Top Partners, but the other Partners might be studied along the same line. Second, our approach might be extended to other resonances, the most obvious candidates being the fermionic Partners of the $2$ light SM quark generations, which are also present in Partial Compositeness. The phenomenology of the latter states is uninteresting for Anarchic Partial Compositeness, and effectively covered by Top Partner searches, but it becomes peculiar and worth studying when Flavor Symmetries are introduced in the model. In the latter case, light generation Partners decay to light SM fermions rather than Top and Bottom and furthermore they can be singly produced with a large rate through their direct coupling with the light quarks in the Proton. First careful assessments of the light partner collider phenomenology was performed in Refs.~\cite{Delaunay:2013pwa,Atre:2008iu,Atre:2011ae} but a systematic interpretation strategy is missing and could be developed following our method. Finally, it could be worth refining our theoretical predictions of the single production rates which, as explained in Section~\ref{sec:prod_mech}, are extracted from available NLO results under some approximation. It should be easy to improve them by complete NLO QCD calculations.

\subsubsection*{Note added}

After this work was completed we became aware of Ref.~\cite{last}, which provides a dedicated analysis for charge $2/3$ ($\widetilde{T}$-like) and $-1/3$ Top Partners singly produced in association with a Bottom quark. The sensitivity of this analysis to the $\widetilde{T}$ single production cross-section is considerably weaker than the one claimed by Ref.~\cite{Ortiz:2014iza}, on which our results are based. This is most likely due to the fact that $2$ $b$-tagged jets are required in Ref.~\cite{last} rather than one as in Ref.~\cite{Ortiz:2014iza}. Given that the second $b$ originates from gluon splitting (see Figure~\ref{fig:sp_gen}), it is preferentially forward and soft and asking for it to be detectable and identifiable costs a considerable price in terms of signal efficiency. Whether or not this second $b$-tag is really needed to reduce the background is an open question, which is important to sort out for a correct assessment of the current $\widetilde{T}$ limits and of the LHC Run-$2$ reach.

\section*{Acknowledgments}

G.P. is grateful to the Mainz Institute for Theoretical Physics (MITP) for its hospitality
and its partial support during the completion of this work. A.W. acknowledges the ERC Advanced Grant no.~267985 {\it DaMeSyFla}, the MIUR-FIRB Grant RBFR12H1MW and the SNF Sinergia no.~CRSII2-141847 for support. The work of O.M. was supported by MIUR under the contract 2010 YJ2NYW-010 and in part by the MIUR-FIRB grant RBFR12H1MW

\appendix

\section{A {\sc MadGraph} model for top partners searches}\label{app:Madgraph_model}

In this section we present a description of the {\sc MadGraph} model designed to simulate the top partners signals.
The model incorporates the resonances which most often appear in the composite Higgs scenarios,
but can be also used to describe any other type of heavy composite fermions interacting predominantly with the third family
of SM quarks. Indeed in the model we keep the couplings of the resonances to the top and bottom quarks as free parameters
and we impose electric charge conservation as the only restriction on the interactions. We do not account for derivative interactions
with a Higgs boson, but they can be brought to a non-derivative form by a suitable field redefinition.
The model is available at \emph{http://hepmdb.soton.ac.uk} under the name ``Simplified Model of Composite Top Partners (STP)''.
The top partners, their charges and the conventions for their couplings are listed in Table~\ref{tab:1}.
\begin{table}
\begin{center}
\begin{tabular}{  c | c | c | c | c | c }
\multicolumn{2}{}{c|} & \multicolumn{4}{|c}{\rule[-.35em]{0pt}{.75em} couplings} \\
\hline
\rule[-.15em]{0pt}{1.25em}partner (MG name) & $Q$  & $W^{\pm}$ & $Z$ & $h$  & $W^{\pm} W^{\pm}$\\
\hline
\rule{0pt}{1.25em}$T_{2/3}$ (T23) &  2/3   & \; $c_{L}^{T W}$,\; $c_{R}^{T W}$\;  & \;$c_{L}^{T Z}$,\; $c_{R}^{T Z}$\; &\; $c_{L}^{T h}$,\; $c_{R}^{T h}$\; & --- \\
\rule{0pt}{1.25em}$B_{1/3}$ (B13) &  -1/3  & \; $c_{L}^{B W}$,\; $c_{R}^{B W}$\;  & \;$c_{L}^{B Z}$,\; $c_{R}^{B Z}$\; &\; $c_{L}^{B h}$,\; $c_{R}^{B h}$\; & --- \\
\rule{0pt}{1.25em}$X_{5/3}$ (X53) &  5/3   & \; $c_{L}^{X W}$,\; $c_{R}^{X W}$\;  & ---  &  ---  &  --- \\
\rule{0pt}{1.25em}$Y_{4/3}$ (Y43) &  -4/3  & \; $c_{L}^{Y W}$,\; $c_{R}^{Y W}$\;  & ---  &  ---  & --- \\
\rule{0pt}{1.25em}$V_{8/3}$ (V83) &  8/3   &  ---  &	---	 & --- & \; $c_{L}^{V W}$,\; $c_{R}^{V W}$\;
\end{tabular}
\end{center}
\caption{List of {\sc MadGraph} model conventions for the top partners names, their electric charges and couplings.}
\label{tab:1}
\end{table}

The couplings $c_{[L/R]}^{[A] [B]}$ are the coefficients in the Lagrangian defining the strength of interaction
of the composite partners with SM top and bottom quarks, up to a factor $g_w/2$ which we introduce explicitly in
case of couplings to gauge bosons. The subscript denotes the chirality of the SM quarks,
while the superscript corresponds to the name of the top partner ($[A]$) and the gauge field or the Higgs boson
involved in the interaction ($[B]$).
The type of SM quark (top or bottom) involved in the vertex follows from the electric charge conservation and is not explicitly
indicated.
For example the $c_{L}^{T W}$, $c_{L}^{T h}$ and $c_{L}^{V W}$ parameters correspond to the following interaction terms in the Lagrangian
\bea
& & {g_w \over 2} c_{L}^{T W} \,  \left[ \overline T_L \, \gamma_{\mu} \, b_L \, W^{\mu} \right] + h.c. \\
& & c_{L}^{T h} \, \left[ \overline T_R \, t_L \, h \right]+ h.c. \\
& & {g_w^2 \over 4} {c_{L}^{V W} \over \Lambda} \, \left[\overline V_R  \, t_L \, W_{\mu} W^{\mu} \right]+ h.c.
\eea
where the dimensionful scale $\Lambda$ (``LAMBDA'' in the MG model with a default value $3$~TeV) appears only in the
couplings of the charge $8/3$ state $V$ (see Ref.~\cite{Matsedonskyi:2014lla} for further details about the $V_{8/3}$ state).
As explained in the main text, in full generality we can assume that all the couplings are real.
In the {\sc MadGraph} model the couplings are given in the format c[L/R][A][B]. The names, allowing to specify the order of the given interaction
needed for the process, are defined as [L/R][A][B] (for instance ``\texttt{generate p p $>$ T b$\sim$ j  LTW=0}'' will only generate processes with
a Right-Handed coupling  $c_R^{TW}$).

Masses and widths are denoted as M[A] and W[A] respectively.
The decay widths are computed automatically for all the partners, except the $V_{8/3}$.
For the $V_{8/3}$ the total width must be set by hand in the model card for each value of the parameters.

\section{Analytic expressions for the decay widths}\label{app:widths}

In this appendix we collect the analytic expressions of the partial widths for the decays
of a fermionic resonance into a SM quark and a gauge field or the Higgs.
These expressions can be easily used to express the resonances branching fractions as
analytical functions of the single production couplings.

The partial width for the decay into a gauge boson $V$ and a SM quark $q$ is given by
\begin{equation}
\Gamma_V = \frac{g_w^2}{32 \pi} \frac{p(M_X, m_q, m_V)}{M_X^2} \left[\left(c_L^2 + c_R^2\right)
\left(
\frac{M_X^2 + m_q^2}{2} + \frac{(M_X^2 - m_q^2)^2}{2 m_V^2} - m_V^2\right)
- 6\, c_L\, c_R\, M_X m_q\right]\,,
\end{equation}
where $M_X$, $m_q$ and $m_V$ are the masses of the heavy resonance $X$, of the SM quark and
of the gauge boson respectively. For shortness we denote by $c_{L,R}$ the $V$-mediated
couplings of the $X$ resonance to the Left- and Right-Handed components of $q$ (these couplings
are denoted by $c_{L,R}^{Vq}$ in the main text). The $p(M_X, m_q, m_V)$ function denotes the
size of the spatial momentum of the final particles in the heavy-resonance rest frame and is given by
\begin{equation}
p(M_X, m_1, m_2) = \frac{\sqrt{\left[M_X^2 - (m_1 + m_2)^2\right]
\left[M_X^2 - (m_1 - m_2)^2\right]}}{2 M_X}\,.
\end{equation}

The partial width for the decay into the Higgs and a SM quark $q$ is given by
\begin{equation}
\Gamma_h = \frac{1}{8 \pi} \frac{p(M_X, m_q, m_h)}{M_X^2} \left[\left(c_L^2 + c_R^2\right)
\frac{M_X^2 + m_q^2 -m_h^2}{2}
+ c_L\, c_R\, M_X m_q\right]\,,
\end{equation}
where $m_h$ denotes the Higgs mass.


\end{document}